\documentclass[12pt]{article}

\usepackage[utf8]{inputenc}

\usepackage[square,numbers,merge,comma,sort&compress]{natbib} 
\makeatletter
\def\NAT@spacechar{\,}
\makeatother

\usepackage{geometry}
\usepackage{mathtools,mathrsfs}
\usepackage{dsfont,relsize}
\usepackage{newtxtext,newtxmath}
\usepackage{amsmath}
\usepackage{amsbsy}
\usepackage{graphicx}
\usepackage{float}

\usepackage[labelsep=colon]{caption}  
\usepackage[labelsep=colon,aboveskip=10pt,belowskip=10pt]{subcaption}
\usepackage{authblk}

\usepackage{xargs}
\newcommandx{\overbar}[1]{\mkern 5.mu\overline{\mkern-5.mu#1\mkern-1.0mu}\mkern 1.5mu}
\DeclareMathAlphabet{\mathpzc}{OT1}{pzc}{m}{it}

\newcommand{\SO}{\textrm{SO}}
\usepackage{xcolor}
\definecolor{Green}{rgb}{0.05, 0.45, 0.25}
\usepackage[breaklinks=true,backref=page]{hyperref}
\hypersetup{
    bookmarks=true,         
    bookmarksnumbered=true,
    pdfmenubar=true,        
    pdffitwindow=false,     
    pdfpagemode={UseNone},
    pdfstartview={FitH},
    colorlinks=false,
    plainpages,
    a4paper,
    linktoc=page,
    urlcolor=Green,
}
\AtBeginDocument{\hypersetup{pdfborder={0 0 1}}}%

\usepackage{amssymb}
\usepackage{upgreek}
\numberwithin{equation}{section}
\allowdisplaybreaks
\usepackage[all]{xy}
\usepackage{color}
\usepackage{graphicx}
\graphicspath{{images/}}

\usepackage[utf8]{inputenc}
\usepackage{ragged2e}
\usepackage{appendix}
\usepackage{slashed}

\title{\LARGE\bf  $\mathcal{N}$-Extended $D=4$ Supergravity, Unconventional SUSY and Graphene
\vskip 1.5cm}

\author[a,b,e]{L.\ Andrianopoli}
\author[a,b,c]{B.L.\ Cerchiai}
\author[a,e]{R.\ D'Auria}
\author[a,b]{A.\ Gallerati}
\author[a,b]{R.~Noris}
\author[a,b,e]{M.\ Trigiante}
\author[d]{J.\ Zanelli}

\affil[a]{Politecnico di Torino, Dip.\ DISAT. Corso Duca degli Abruzzi 24, 10129 Torino, Italy}
\affil[b]{Istituto Nazionale di Fisica Nucleare (INFN) Sezione di Torino, Italy}
\affil[c]{Museo Storico della Fisica e Centro Studi ``Enrico Fermi'', Piazza del Viminale 1, 00184 Roma, Italy}
\affil[d]{Centro de Estudios Cient\'{\i}ficos (CECs), Av. Arturo Prat 514, Valdivia, Chile}
\affil[e]{Arnold-Regge Center, via P. Giuria 1, 10125 Torino, Italy}
\date{}

\begin{document}

\maketitle

\vskip 1cm

\begin{abstract}
\noindent
We derive a $2+1$ dimensional model with unconventional supersymmetry at the boundary of an ${\rm AdS}_4$ $\mathcal{N}$-extended supergravity, generalizing previous results. The (unconventional) extended  supersymmetry of  the boundary model is instrumental in describing, within a top-down approach, the electronic properties of {graphene-like 2D materials} at the two Dirac points, ${\bf K}$ and ${\bf K}'$. The two valleys correspond to the two independent sectors of the ${\rm OSp}(p|2)\times {\rm OSp}(q|2)$ boundary model in the $p=q$ case, which are related by a parity transformation. The Semenoff and Haldane-type masses entering the corresponding Dirac equations are identified with the torsion parameters of the substrate in the model.
\end{abstract}

\vfill
\noindent
{\footnotesize{\tt laura.andrianopoli@polito.it};\\
{\tt bianca.cerchiai@polito.it}; \\
{\tt riccardo.dauria@polito.it}; \\
{\tt antonio.gallerati@polito.it}; \\
{\tt ruggero.noris@polito.it}; \\
{\tt mario.trigiante@polito.it}; \\
{\tt z@cecs.cl}}

\numberwithin{equation}{section}
\clearpage

\tableofcontents

\bigskip

\section{Introduction}
Chern-Simons theories including gravity and matter in three dimensions developed three decades ago by Ach\'ucarro and Townsend \cite{Achucarro:1987vz}, have been shown to exhibit interesting features \cite{Gaiotto:2008sd,Kapustin:2009cd}, particularly in connection with the holographic correspondence \cite{Maldacena:1997re,Gubser:1998bc,Witten:1998qj}. Our interest here focuses on the Ach\'ucarro-Townsend (AT) theory, following the Ansatz proposed in \cite{Alvarez:2011gd,Guevara:2016rbl} (referred to as the AVZ model in the sequel). The AVZ model consists of a Chern-Simons system in $2+1$ dimensions for the supergroup ${\rm OSp}(2|2)$. It is an effective theory for a massive spin-1/2 fermion, generically defined on a curved geometry and minimally coupled to the background gravity and a $\textrm{U}(1)$ gauge field. This system exhibits an unconventional form of supersymmetry based on a graded Lie algebra that extends the local invariance of the tangent of the spacetime manifold, with the addition of the internal gauge generators necessary to close the superalgebra. All the fields are contained in the gauge connection for the adjoint representation of the supergroup, namely%
\footnote{Here we assume $\mathbb{Q}$ and $\psi$ to have dimensions $({\rm length})^{-\frac{1}{2}}$ and $({\rm length})^{\frac{1}{2}}$, respectively, as it is common in the supergravity literature. The one-form gauge fields $A$ and $\omega$ on the other hand, are assumed to be dimensionless.}
\begin{equation} \label{A}
\mathbb{A}= \frac{1}{2}\,\omega^{ij}\, \mathbb{J}_{ij} + A\cdot \mathbb{T} + \overline{\psi}_A\,\mathbb{Q}^A + \overline{\mathbb{Q}}_A\,\psi^A + \dots\;,
\end{equation}
where $\omega^{ij}$, $A$ and $\psi^A$ are one-forms, while ${\rm \mathbb{J}}$, $\mathbb{T}$ and $\mathbb{Q}$ are the generators of Lorentz, internal gauge and supersymmetry transformations, respectively. What makes this model unconventional is that it assumes a peculiar Ansatz for the fermionic gauge fields $\psi^A_{\mu}$, expressing them as composite fields of the vielbein $e^i$ and spin-$1/2$ fields $\chi_A$,
\begin{align}\label{Ansatzzanelli}
\psi^A=i\,\gamma_i\,e^i \chi^A.
\end{align}
In three dimensions, the simplest Lagrangian for the connection (\ref{A}) is a Chern-Simons form and the resulting AVZ model is particularly suited for describing graphene near the Dirac points in a generic spatial lattice with nonvanishing curvature and torsion \cite{Iorio:2011yz}.

In $3+1$ dimensions, on the other hand, the simplest action for a superconnection of the form (\ref{A}) is a Yang-Mills theory for the smallest superalgebra that extends the AdS$_4$ symmetry and yields a spin-1/2 field minimally coupled to Einstein gravity and the Maxwell field \cite{Alvarez:2013tga}. This four-dimensional unconventional SUSY model was in turn shown to correspond to the boundary theory of a Chern-Simons theory for a super connection in five-dimensions \cite{Gomes:2017rmd}.

In what follows, we consider a three dimensional model of unconventional supersymmetry at the boundary of an AdS$_4$ supergravity vacuum, extending the analysis in \cite{Andrianopoli:2018ymh}, where it was shown that the three-dimensional AVZ model could be holographically realized as the boundary theory of an $\mathcal{N}=2$ four-dimensional supergravity of the AdS$_4$ spacetime. The model in \cite{Andrianopoli:2018ymh}  was constructed by embedding the ${\rm OSp}(2|2)$--Chern-Simons theory of the AVZ model in an $\mathcal{N}=2$, $D=3$ AdS$_3$ supergravity described by an AT theory with gauge group ${\rm OSp}(2|2)_+\times {\rm SO}(1,2)_-$. This theory was in turn obtained as an effective model at the boundary of an AdS$_4$ space on which an $\mathcal{N}=2$ supergravity is defined. This results from a suitable choice of boundary conditions for the four-dimensional fields. Imposing then the AVZ Ansatz (\ref{Ansatzzanelli}) for the $D=3$ fermions identifies the resulting spin-$1/2$ fields $\chi_A$ as the radial component of the four-dimensional gravitini whose mass is related to the AdS$_3$ radius. Applying the resulting model to the effective description of the electronic properties of graphene {and other graphene-like 2D materials\footnote{{By \emph{graphene-like} materials we mean two-dimensional materials featuring a honeycomb lattice and an emergent behaviour as Dirac fermions for the pseudo-particle wavefunction.}}} provides a top-down approach to the understanding of the origin of  supersymmetric phenomenology of this physical system \cite{Ezawa:2006cr,Lee:2006if,Dartora:2013psa}. The quantum BRST formulation of the same $D=3$ model was discussed in \cite{Andrianopoli:2019sqe}.

The aim of the present paper is to generalize  the construction of \cite{Andrianopoli:2018ymh} to an $\mathcal{N}$-extended supergravity, with maximally supersymmetric AdS$_4$ vacuum related to an ${\rm OSp}(p|2)_+\times {\rm OSp}(q|2)_-$, $p+q=\mathcal{N}$, Chern-Simons theory at the boundary.

It is achieved by generalizing the boundary conditions for the $D=4$ fields used in  \cite{Andrianopoli:2018ymh}. Applying the AVZ Ansatz for
for $\psi_{A\,\mu}$ ($A=1,\dots, \mathcal{N}$) an effective model for the massive spin-$1/2$ fields $\chi_A$ on a curved background in the presence of a larger amount of supersymmetry and a larger internal symmetry group ${\rm SO}(p)\times {\rm SO}(q)$ is obtained. This allows to introduce extra internal degrees of freedom which can provide an application of the model to the description of graphene. The supersymmetry of the boundary model is defined by the partition $(p,q)$ of $\mathcal{N}$ and depends on the signature of a real symmetric matrix $\eta_{AB}$ entering the boundary conditions for the gravitini.

Besides discussing the relation to the four-dimensional supergravity, the effective theory for the $\chi_A$ fields is explicitly constructed and its symmetries illustrated. The fermionic fields $\chi_A$ naturally split into two sets, $\chi_{a_1}$ and $\chi_{a_2\,}$, with $a_1=1,\dots, p$ and $a_2=1,\dots, q$, which correspond to the representations ${\bf \left(\frac{1}{2},0\right)}$ and ${\bf \left(0,\frac{1}{2}\right)}$ of the AdS$_3$ isometry group ${\rm SL}(2,\mathbb{R})_+\times {\rm SL}(2,\mathbb{R})_-$.
In the special case $p=q$, a manifest \emph{reflection symmetry} emerges in the model, under which the fermions in the two sets are interchanged. In light of this we argue that, in this particular case, the Dirac fermions $\chi_{a_1},\, \chi_{a_2}$ may possibly describe the wave functions of the $\pi$-electrons in {graphene-like systems} at the two inequivalent Dirac points ${\bf K},\,{\bf K}'$. Both fermions have masses which, as in the original AVZ model, depend on the torsion of the three-dimensional spacetime. We briefly elaborate on a microscopic (i.e.\ at scales comparable with the honeycomb lattice spacing) description of {graphene-like materials} which can account for the massive Dirac equations that we find at the two Dirac points.

We emphasize that our construction follows a top-down approach, in that the effective $D=3$ theory that we derive at the boundary of AdS$_4$ originates from a well defined supersymmetric effective supergravity in the bulk. Nevertheless it bears similarities with the $D=3$ models considered in \cite{Iorio:2018agc}, whose formulation also features an underlying AdS$_3$ symmetry. We shall elaborate on this at the end of section \ref{4}.\par\smallskip

The paper is organized as follows:

In Sect.\ \ref{2} we discuss the emergence of an $\mathcal{N}$-extended Ach\'ucarro-Twonsend model in $D=3$ at the boundary of an ${\rm AdS}_4$ supergravity by choosing appropriate boundary conditions on the supergravity fields. We also discuss the effect of a reflection transformation on the boundary model, showing that in the $p=q$ case it becomes a symmetry.

In Sect.\ \ref{3} we implement the AVZ Ansatz (\ref{Ansatzzanelli}) in the $\mathcal{N}$-extended AT theory of Sect.\ \ref{2} and discuss the properties of the resulting model and its symmetries in the presence of a general world-volume torsion. In particular, we write the Dirac equations for the spin-$1/2$ fields $\chi_\pm$ in the two sectors acted on by each simple factor of the supergroup, and relate the corresponding masses to the torsion parameters.

In Sect.\ \ref{4} we apply our model as an effective long wave-length description of the electronic properties of {graphene-like Dirac materials}. We show that in the particular case $p=q$ the spin-$1/2$ fields $\chi_\pm$ can be consistently related to the electron wave-functions in two ${\bf K}$ and ${\bf K}'$ Dirac points. This allows us to identify the parity-even and odd components of the corresponding masses with Semenoff and Haldane-type mass contributions, respectively. These quantities, in light of the discussion in Sect.\ \ref{3}, are then consistently expressed in terms of the torsion parameters of the model. In the last subsection we consider a different model within our general construction, defined by $p=4$ and $q=0$, which allows us to make contact with the analysis in \cite{Iorio:2018agc}.

We end with a general discussion of our results and possible generalizations thereof. The appendices contain our notations and conventions and a brief account of some facts about {graphene-like systems} which are relevant to our analysis.

\section{Ach\'ucarro-Townsend \texorpdfstring{$D=3$}{D=3} Theory from \texorpdfstring{AdS$_4$}{AdS4} Supergravity
}\label{2}
The starting point of our analysis is an AdS$_4$ vacuum of an $\mathcal{N}$-extended pure supergravity theory preserving all $\mathcal{N}$ supersymmetries. The vacuum symmetry is described by the supergroup $\mathrm{OSp}(\mathcal{N}|4)$ group%
\footnote{We work with the ``mostly minus'' convention for the metric and refer to a parametrization of AdS$_4$ given by $ds^2=\frac{r^2}{\ell^2}\,ds^2_{{}_{(3)}}-\ell^2\,\frac{dr^2}{r^2}$, where $ds^2_{{}_{(3)}}$ is a locally AdS$_3$ metric of the conformal UV boundary at $r\rightarrow \infty$.}. We consider fluctuations on this background which exhibit the full vacuum symmetry at radial infinity. We require, in particular, all scalar and spin-$1/2$ fields at the conformal boundary to be frozen at their vacuum values, and that the remaining fields obey the $\mathfrak{osp}(\mathcal{N}|4)$ Maurer-Cartan equations. This condition is satisfied, for instance, by the four dimensional vacuum configuration. We therefore do not consider here boundary terms depending on the scalar fields of the $\mathcal{N}$-extended AdS$_4$-supergravity. For pure $\mathcal{N}=2$, where scalar fields are not present, this choice of asymptotic symmetry is consistent with the analysis of \cite{Andrianopoli:2014aqa}, applied in \cite{Andrianopoli:2018ymh}.

The dual description of the $\mathfrak{osp}(\mathcal{N}|4)$ algebra is given in terms of the connection
\begin{align}
\mathbb{A}=\theta^i \otimes E_i=\frac{1}{2}\omega^{\mathcal{A}\mathcal{B}}\,L_{\mathcal{A}\mathcal{B}}+
\frac{1}{2}A^{CD}\,T_{CD}+\overline\Psi_{\alpha}^A\,Q_A^{\alpha},
\end{align}
where $L_{\mathcal{A}\mathcal{B}}$ ($\mathcal{A},\mathcal{B}=0,\dots, 4$) and $T_{AB}$ (${A},{B}=1,\dots, \mathcal{N}$) are the $\SO(2,3)$ and $\SO(\mathcal{N})$ generators respectively, whereas $Q_A^{\alpha}$ ($\alpha=1,\dots,4$) are the Majorana supersymmetry generators (see Appendix \ref{App_A}). The structure of the algebra is encoded in the Maurer-Cartan equations $d\mathbb{A}+\mathbb{A}\wedge \mathbb{A}=0$, which can be conveniently written in a manifestly covariant form with respect to the $D=4$ Lorentz group $\SO(1,3)$ by splitting $\mathcal{A}=(a,4)$, with $a=0,1,2,3$ and defining \,$L_{a4}:={\ell}\,P_a$\,,  \,$\omega^{a4}:=\ell^{-1}\,V^a$, where $V^a$ is the four-dimensional vielbein:
\begin{equation}
\begin{split}
&d\omega^{ab}+\omega^{a}{}_{c}\wedge\omega^{cb}-\ell^{-2}V^a\wedge V^b-\frac{1}{2{\ell}}\left(\overline\Psi^A\wedge\Gamma^{ab}\Psi_{A}\right)=0, \\
&dV^a+\omega^{a}{}_{b}\wedge V^b-\frac{i}{2}\left(\overline\Psi^A\wedge\Gamma^a\Psi_A\right)=0, \\
&dA^{CD}+A^C{}_B\wedge A^{BD}+\ell^{-1}\left(\overline\Psi{}^C\wedge\Psi^D\right)=0, \\
&d\Psi^A+\frac{1}{4}\,
\omega^{ab}\wedge\Gamma_{ab}\Psi^A+\frac{i}{2{\ell}}\,
V^a\wedge\Gamma_a\Psi^A+A^{AB}\wedge
\Psi^B=0.\label{MCADS}
\end{split}
\end{equation}

As mentioned above, we shall restrict to $D=4$ asymptotically anti-de Sitter backgrounds in which the above algebra \emph{only} holds at the UV boundary located at radial infinity.

To proceed, it is convenient to rewrite the Maurer-Cartan equations in a form which is covariant with respect to the Lorentz group at the spatial boundary, $\SO(1,2)$. This is achieved by further splitting the rigid index $a$ into $a=(i,3)$, where $i=0,1,2$ labels the boundary dreibein and $a=3$ labels the vierbein along the radial direction. This allows to decompose the AdS$_4$ superalgebra in terms of $\SO(1,1)\times \SO(1,2)\subset \SO(2,3)$ where $L_{34}$ and $L_{ij}$ are the $\SO(1,1)$ and $\SO(1,2)$  generators, respectively. We then write the asymptotic algebra in a way in which the $\SO(1,1)$-grading of the fields is manifest. This is achieved by defining
\begin{align}
E^i_\pm &:=\pm\frac{1}{2}\left(V^i\mp \ell\omega^{3i}\right)\;,
\\
\omega^{i3} & =\frac{1}{\ell}\,(E^i_+ + E^i_-)\;,
\\
V^i &= E^i_+ - E^i_- \;,
\end{align}
and decomposing the gravitini in their chiral components with respect to the same $\SO(1,1)$, represented on the spinors by the matrix $\Gamma^3$,
\begin{align}
\Psi^{A\alpha}=\Psi^A_++\Psi^A_-\;,\qquad
\Gamma^3\Psi^A_\pm=\pm i\,\Psi^A_\pm\,.\quad
\end{align}
With these definitions, the previously obtained asymptotic relations  become%
\footnote{We use $\overline\Psi_\pm\Gamma^{ij}\Psi_\pm=\overline\Psi_\pm\Gamma^{3}\Psi_\pm=\overline\Psi_\pm\Psi_\pm=0$.}
\begin{equation} 
\begin{split}
&\left.d\omega^{ij}+\omega^{i}{}_{k}\wedge\omega^{kj}+\frac{4}{{\ell}^2}\,E_+^{[i}\wedge E_-^{j]}-\frac{1}{{\ell}}\left(\overline\Psi^A_+\wedge\Gamma^{ij}\Psi_{A-}\right)\right\vert_{\partial \mathcal{M}}\!=0\,,
\\
&\left.dE_\pm^i+\omega^i{}_j\wedge E_\pm^j\mp\frac{1}{{\ell}}\,E^i_\pm\wedge V^3\mp\frac{i}{2}\left(\overline\Psi^A_\pm\wedge\Gamma^i\Psi_{A\,\pm}\right)\right\vert_{\partial \mathcal{M}} \!=0\,,
\\
&\left.dV^3-\frac{1}{\ell}\,(E^i_+ + E^i_-)\wedge V_i+\overline\Psi^A_-\wedge\Psi_{A+}\right\vert_{\partial \mathcal{M}}\!=0\,,
\\
&\left.dA^{CD}+A^C{}_M\wedge A^{MD}+\frac{2}{{\ell}}\left(\overline\Psi{}^{[C}_+\wedge\Psi^{D]}_-\right)\right\vert_{\partial \mathcal{M}}\!=0\,,
\\
&\left.d\Psi^{M\beta}_\pm+\frac{1}{4}\,\omega^{ij}\wedge\left(\Gamma_{ij}\Psi^M_\pm\right)^\beta\pm
\frac{i}{{\ell}}\,E_\pm^i\wedge\left(\Gamma_i\Psi^{M}_\mp\right)^\beta \pm\frac{1}{2{\ell}}\,V^3\wedge\Psi^{M\beta}_\pm +\delta^M{}_{[C}\delta_{D]B}\,A^{CD}\wedge\Psi^{B\beta}_\pm\right\vert_{\partial \mathcal{M}}\!=0\,.
\label{MCboundary}
\end{split}
\end{equation}

\subsection{Boundary limit}
We are now interested into the AdS$_4$ boundary, which is reached in the limit $r\to\infty$. In order to perform the limit, we define the vielbein at the boundary as\footnote{Note the more symmetric choice of the numerical factors entering the boundary conditions in the  ``$+$'' and ``$-$'' sectors with respect to  \cite{Andrianopoli:2018ymh}. This is achieved by a rescaling $r\rightarrow r/2$, which does not alter the asymptotic limit.}

\begin{align}
E^i_+(r,x)&=\frac 12\left(\frac{r}{{\ell}}\right)E^i(x)+
O\left(\frac{{\ell}^2}{r^2}\right)\,,
\qquad
E^i_-(r,x)=-\frac{1}{2}\left(\frac{{\ell}}{r}\right)E^i(x)+O\left(\frac{{\ell}^2}{r^2}\right)\,,
\end{align}
where $x=(x^\mu)$, $\mu=0,1,2$,  are the boundary coordinates. As for the spinors we require
\begin{align}
\Psi^A_{+\,\mu}(r,x)\,dx^\mu &=
\sqrt\frac{r}{2\ell}\begin{pmatrix*}[c]
  \psi^A(x)  \\
  \bf 0
\end{pmatrix*}+O\left(\frac{{\ell}}{r}\right), \\
\Psi^A_{-\mu}(r,x)\,dx^\mu &=
\sqrt\frac{{\ell}}{2 r}\begin{pmatrix*}[c]
\bf 0  \\
  \eta^{AB}\psi_B(x)
  \end{pmatrix*}+O\left(\frac{{\ell}}{r}\right),
\end{align}
where $\eta_{AB}$ is a symmetric metric such that $\eta_{AC}\, \eta_{CB}=\delta_{AB}$\footnote{Being $A,B$ indices in  the fundamental $\mathcal N$-dimensional representation of the R-symmetry group $\SO(\mathcal N)$, we do not distinguish between their upper and lower positions.}. As far as the spin and the gauge connection, we require
\begin{align}
\omega^{ij}(r,x)=\omega^{ij}_\mu(x)\,dx^\mu+\dots\,,\qquad
A^{AB}(r,x)=A^{AB}_\mu(x)\,dx^\mu+\dots\,,
\end{align}
where the ellipses denote subleading terms in ${\ell}/r$. When writing the equations at the boundary we shall simply denote by $\omega^{ij}$ and $A^{AB}$ the boundary values of the corresponding quantities in the bulk, as defined by the leading terms above.

Consistency of the boundary conditions requires that both $V^3$ and $dV^3$ vanish at the boundary. This is indeed the case since, at the boundary, $\omega^3{}_i\wedge E^i=0$ (by virtue of the general properties of the extrinsic curvature of the boundary) and $(\overline{\Psi_\mp}) \Psi_\pm \propto \eta_{AB}\,({\psi}^A)^T \sigma^2 \psi^B=0$, the latter being a consequence of the antisymetry of $\sigma^2$ and the symmetry of $\eta_{AB}$.

With these boundary conditions, one can verify that the {boundary equations \eqref{MCboundary}} involve only boundary fields%
\footnote{More specifically, it is straightforward to verify that the right-hand-side of equations (\ref{MCboundary}) are proportional, through powers of ${\ell}/r$, to the right hand side of equations (\ref{curvatures2}).}%
\begin{equation}\label{curvatures2}
\begin{split}
&d\omega^{ij}+\omega^{i}{}_{k}\wedge\omega^{kj}-\frac{1}{{\ell}^2}\,E^i\wedge E^j-\frac{1}{2{\ell}}\,\left(\overline\psi^A\wedge\gamma^{ij}\eta_{AB}\psi^{B}\right)=0\,, \\
&dE^i+\omega^i{}_j\wedge E^j-\frac{i}{2}\,\left(\overline\psi^A\wedge\gamma^i\psi_A\right)=0\,, \\
&dA^{CD}+A^C_M\wedge A^{MD}+\frac{1}{{\ell}}\,\overline\psi{}^{[C}\wedge\eta^{D]B}\psi_{B}=0\,,
\\
&d\psi^{A}+\frac{1}{4}\,\omega^{ij}\,\wedge\gamma_{ij}\,\psi^A+\frac{i}{2{\ell}}\,E^i\wedge
\gamma_i\,\eta^{AB}\psi_{B} +A^{AB}\wedge\psi^{B}=0\,.
\end{split}
\end{equation}
\sloppy
We now examine the role of the $\eta_{AB}$ in the breaking of the $D=4$ R-symmetry group: ${{\rm O}(\mathcal{N})\rightarrow {\rm O}(p)\times {\rm O}(q)}$, $p+q=\mathcal{N}$, where the integers $p,\,q$,  define the signature of $\eta$. Indeed, through an ${\rm O}(\mathcal{N})$ rotation $\eta$ can be brought to the diagonal form
\begin{align}\label{etadiag}
\eta_{AB}=\begin{pmatrix*}
\textbf{1}_{ p \times p} & \textbf{0}_{p\times q} \\
\textbf{0}_{q\times p} & -\textbf{1}_{ q \times q}
\end{pmatrix*}\,.
\end{align}
\sloppy
The index $A=1,\dots, \mathcal{N}$ then naturally splits into $A=(a_1,a_2)$, where $a_1=1\,\dots,\,p $ and ${a_2=p+1\,\dots,\,\mathcal{N}}$, and similarly $B=(b_1,\,b_2)$. Inserting expression (\ref{etadiag}) into (\ref{curvatures2}) one can verify that, in the Maurer-Cartan equations for the gauge fields $A^{BC}$, the fermion bilinear is projected on the adjoint of the algebra generating the subgroup ${\rm O}(p)\times {\rm O}(q)$ of ${\rm O}(\mathcal{N})$, so that the equation for the gauge fields $A^{a_1b_2}$, associated with the generators in the coset ${\rm O}(\mathcal{N})/({\rm O}(p)\times {\rm O}(q))$, reads
\begin{equation}
dA^{a_1b_2}+A^{a_1}{}_{c_1}\wedge A^{c_1b_2}+A^{a_1}{}_{c_2}\wedge A^{c_2b_2}=0\,,
\end{equation}
and therefore these fields can be consistently set to zero: $A^{a_1b_2}=0$. This condition, however, is not optional, implicit in the requirement that the fields at the boundary (and in particular the gravitini fields) satisfy consistent equations.  The remaining equations describe the superalgebra of ${\rm OSp}(p|2)_+\times {\rm OSp}(q|2)_-$ (note that the subscripts ``$\pm$'', from now on, no longer refer to
the eigenvalues of $-i\Gamma^3$, but to the two factors in the $D=3$ supergroup). To see this we  define
\begin{equation}
\begin{split}
&\Omega_{(\pm)}^i:=\,\omega^i\pm\frac{E^i}{{\ell}}\;,\quad
\psi_+:=(\psi^{a_1})\,\,,\,\,\,\,\,\psi_-:=(\psi^{a_2})\;,\quad
A_+:=(A^{a_1b_1})\;,\quad
A_-:=(A^{a_2b_2})\;,\\
&\mathcal{D}[\Omega_+,\,A_+]\psi_+ :=\left(d\psi^{a_1}+\frac{i}{2}\Omega^i_+\wedge\gamma_i\psi^{a_1}+
A^{a_1b_1}\wedge\psi^\beta_{b_1}\right)\;,\\
&\mathcal{D}[\Omega_-,\,A_-]\psi_- :=\left(d\psi^{a_2}+\frac{i}{2}\Omega^i_-\wedge\gamma_i\psi^{a_2}+
A^{a_2b_2}\wedge\psi_{b_2}\right)\;,
\end{split}
\end{equation}
where $\omega^i:=\,\frac{1}{2}\,\epsilon^{ijk}\,\omega_{jk}$. Equations (\ref{curvatures2}) can then be cast into the following compact form
\begin{subequations}\label{AT-likecurv}
\begin{align}
&R_\pm {}^i:=d\Omega_\pm^i-\frac{1}{2}\epsilon^{ijk}\Omega_{\pm \,j} \wedge\Omega_{\pm\,k}=\pm\frac{i}{{\ell}}\left(\overline\psi_\pm\wedge\gamma^{i}\psi_\pm\right)\,;
\\[\jot]
&\mathcal{D}[\Omega_\pm,\,A_\pm]\psi_\pm=0\,,\label{eq:DOmAPsi}
\\[\jot]
&\mathcal{F}^{a_1 b_1}:=dA^{a_1b_1}+A^{a_1}{}_{c_1}\wedge A^{c_1b_1}=-\frac{1}{{\ell}}\,\left(\overline\psi^{a_1}\wedge\psi^{b_1}\right)\,,
\\
&\mathcal{F}^{a_2 b_2}:=
dA^{a_2b_2}+A^{a_2}{}_{c_2}\wedge A^{c_2b_2}=\frac{1}{{\ell}}\left(\overline\psi^{a_2}\wedge\psi^{b_2}\right)\,,
\end{align}
\end{subequations}
which rflects the structure of the $\mathfrak{osp}(p|2)_+\oplus\mathfrak{osp}(q|2)_-$  superalgebra%
\footnote{Note that, in our conventions, the structure constants of an $\mathfrak{sl}(2,\mathbb{R})$ algebra are chosen to be $f_{ijk}=-\epsilon_{ijk}$.}, where the two parts are described by the 1-forms \,$\Omega^i_+,\,\psi_+,\,A_+$\, and \,$\Omega^i_-,\,\psi_-,\,A_-$, respectively\footnote{These curvatures are exactly those obtained by Ach\'ucarro and Townsend in \cite{Achucarro:1987vz} in three dimensions starting from a Chern-Simons action, if we call
\;\;$\psi^{i,i'}\to\sqrt{2}\,\psi^{a_1,a_2}, \;\;\; \overline\psi {}^{i,i'}\to i\,\sqrt{2}\,\overline\psi {}^{a_1,a_2}, \;\;\;   m=\frac{1}{2{\ell}} \,.$}.

The Maurer-Cartan equations (\ref{AT-likecurv}) can be derived from the Lagrangian 3-form \cite{Achucarro:1987vz}
\begin{equation}\label{Lagrangianpm}
\begin{split}
\mathcal{L}=\;&\mathcal{L}_{(+)}-\mathcal{L}_{(-)}-\frac{1}{2}\,d(\Omega_+{}_k\wedge \Omega^k_-)\,,\\[1.5\jot]
\mathcal{L}_{(\pm)}:=\;&\frac{1}{2}\,\left(\Omega_{\pm\,i}d\Omega_\pm^i-\frac{1}{3}\,
\epsilon_{ijk}\,\Omega_\pm^i\wedge \Omega_\pm^j\wedge \Omega_\pm^k \right)+\,{\rm Tr}\left(A_\pm\wedge dA_\pm+\frac{2}{3}\,A_\pm\wedge A_\pm\wedge A_\pm\right)\pm\\
&\pm \frac{2}{\ell}
\overline{\psi}_\pm\wedge\mathcal{D}[\Omega_\pm,\,A_\pm]\psi_\pm\,,
\end{split}
\end{equation}
where the total derivative \;$-\frac{{\ell}}{2}\,d(\Omega_+{}_k\wedge \Omega^k_-)=d(\omega_i\wedge E^i)$\; is a Gibbons-Hawking term which originates from writing the Einstein-Hilbert Lagrangian as the difference of two Chern-Simons forms. Note that one could have chosen a singular matrix $\eta_{AB}$, with $p$ positive, $q$ negative and $r$ vanishing eigenvalues ($\mathcal{N}=p+q+r$). In that case, $\eta^2=\mathbb{P}_{p+q}=({\bf 1}-\mathbb{P}_r)$, where $\mathbb{P}_r$ is the projector on the 0-eigenspace of $\eta_{AB}$. Imposing \,${\mathbb{P}_r}^A{}_B\,A^{BC}=0$,\, \,${\mathbb{P}_r}^A{}_B\,\psi^B=0$\, and \,$A^{a_1b_2}=0$\, at the boundary, the resulting boundary fields are connections in the algebra $\mathfrak{osp}(p|2)_+\oplus\mathfrak{osp}(q|2)_-$, corresponding to the smaller supersymmetry $\mathcal{N}'=p+q=\mathcal{N}-r$. This more general choice of $\eta_{AB}$ therefore allows a reduced amount of supersymmetry at the boundary.

\subsection{Reflection transformations and the symmetric case \texorpdfstring{$p=q$}{p=q} }\label{parity}
In this subsection we discuss the effect of a parity transformation on the AT model. This transformation can be characterized as a spatial reflection in the Y-axis tangent to the the 2+1 dimensional boundary ($t\rightarrow t,\; x\rightarrow -x,\; y\rightarrow y$), and implemented on three-dimensional vectors by the matrix $\mathcal{O}_Y={\rm diag}(+1,-1,+1)$.
Recalling that $E^i$ are vectors and $\omega^i$ pseudo-vectors, the transformation properties under this parity of the $E^i$ and $\omega^i$ fields are:
\begin{align}
E^i\rightarrow \tilde{E}^{ i}=\mathcal{O}_Y{}^{i}{}_{j}\,E^j\,,
\qquad
\omega^i\rightarrow \tilde{\omega}^{ i}=-\mathcal{O}_Y{}^{i}{}_{j}\,\omega^j\,,\label{parity1}
\end{align}
\sloppy
which in turn implies: $\Omega^i_\pm\rightarrow \tilde{\Omega}^{i}_\pm=-\mathcal{O}_Y{}^{i}{}_{j}\,\Omega^j_\mp$\,.
The action on the full supersymmetric model is not in general an invariance, since the ${\rm OSp}(p|2)_+\times {\rm OSp}(q|2)_-$ model is mapped into the ${{\rm OSp}(q|2)_+\times {\rm OSp}(p|2)_-}$ one, i.e.\ the + and - sectors are interchanged.\par
In the special case $p=q$ this discrete  transformation is an invariance of the theory.
To make the parity symmetry manifest in the supersymmetric case we extend its action to the fermionic\footnote{The Clifford algebra representation that we use is given in Appendix \ref{App_A}.} and gauge sectors as follows:
\begin{align}
\psi_\pm\rightarrow \tilde{\psi}_\pm=\sigma^1\,\psi_\mp\,,
\quad\;\;
A_\pm\rightarrow \tilde{A}_\pm=A_\mp\,.
\label{parity2}
\end{align}
The reader can easily verify, given our spinor conventions, that:
\begin{equation}
\begin{split}
{\overline{{\psi}}_\pm\wedge\gamma^i{\psi}_\pm}&\,\rightarrow\,
\mathcal{O}_Y{}^{i}{}_{j}\;\overline{\psi}_\mp\wedge
 \gamma^j\psi_\mp\,,
\\[\jot]
{\Omega}_\pm\,\gamma^i{\psi}_\pm&\,\rightarrow\,\Omega_\mp\;\sigma^1\gamma^i\psi_\mp\,,
\\[\jot]
\overline{{\psi}}_\pm\,\mathcal{D}[{\Omega}_\pm\,,\,{A}_\pm]\,{\psi}_\pm &\,\rightarrow\, -\overline{\psi}_\mp\,\mathcal{D}[\Omega_\mp\,,\,A_\mp]\,\psi_\mp\,.
\end{split}
\end{equation}
and that the Maurer-Cartan equations (\ref{AT-likecurv}) are invariant for $p=q$.
The reader can show  that $\tilde{\mathcal{L}}_{(\pm)}=\mathcal{L}_{(\mp)}$, so that the Lagrangian density $\mathcal{L}$ is odd: $\tilde{\mathcal{L}}=-\mathcal{L}$. When making contact with the effective description of {certain 2D materials}, in the next sections, we will, in the $p=q$ case, interpret the $+$ and $-$ sectors as related to the Dirac points ${\bf K},\,{\bf K}'$. \par
As pointed out above, in the general $p\neq q$ case, the discrete reflection symmetry exchanges the $+$ and the $-$ sector so that a $(p,q)$ model is mapped into the $(q,p)$ one.\par
By the same token one can show that the inversion in the X-axis, tangent to the $2+1$ dimensional world volume, is also a symmetry of the model for $p=q$. Its action on three-dimensional vectors is implemented by the matrix $\mathcal{O}_X={\rm diag}(+1,+1,-1)$, while on the gravitini it involves multiplication by the Pauli matrix $\sigma^3$: \,$\psi_\pm\rightarrow \tilde{\psi}_\pm=\sigma^3 \psi_\mp$\,.

\section{Generalized AVZ model}\label{3}
\sloppy
{The Chern-Simons theory discussed in \cite{Alvarez:2011gd} and \cite{Andrianopoli:2018ymh} is naturally defined on a principal fiber bundle $[\mathcal{M}_3,\rm{OSp}(2|2)]$, under the assumption that the bosonic subgroup ${\mathrm{SO}(1,2)\subset\mathrm{OSp}(2|2)}$ of the fiber gauge group is identified with the Lorentz group on the tangent space of the three-dimensional world-volume space time  $\mathcal{M}_3$. This identification  is implicit, in particular, in the Ansatz (\ref{Ansatzzanelli}), where {the $\gamma$-matrices act both on the odd generator $Q_A^\alpha$ of the $\mathrm{OSp}(2|2)$ gauge group and on the world-volume spinor $\chi^\alpha$}. A more general point of view was adopted in \cite{Andrianopoli:2019sqe}, where such identification is not imposed a priori. This allows in particular the construction of the quantum world-volume field theory, in the spirit of the holographic AdS/CFT correspondence \cite{Gaiotto:2008sd,Kapustin:2009cd}.}

\sloppy
{In this paper we are going to analyze some applications to {graphene-like systems} of the geometrical features of the classical model of \cite{Alvarez:2011gd,Andrianopoli:2018ymh} in the more general case of a gauge group ${{\rm OSp}(p|2)_+\times {\rm OSp}(q|2)_-}$. However we prefer to maintain  the conceptual distinction between target space and world volume.} From this point of view, $\Omega_\pm,\,\psi_\pm, \,A_\pm$ are world-volume gauge fields with values in the $\mathfrak{osp}(p|2)_+\oplus\mathfrak{osp}(q|2)_-$ superalgebra (target space). We assume the three-dimensional world-volume to have a tangent bundle with local AdS$_3$ symmetry of radius $\ell'$, and a local frame bundle defined by the dreibein $e^i$. Supersymmetry is just the odd part of the gauge supergroup and is not assumed to act on the world volume at the classical level.

{According to this more general point of view,} the isometry group ${\rm SL}(2,\mathbb{R})'_+\times {\rm SL}(2,\mathbb{R})'_-$ of the tangent {space} to the {world-volume} geometry and the bosonic subgroup ${\rm SL}(2,\mathbb{R})_+\times {\rm SL}(2,\mathbb{R})_-$ of the gauge group are in principle unrelated, and we shall use primed and unprimed symbols to emphasize this distinction. In particular, the connections $\Omega^{\prime\,i}_\pm$ of ${\rm SL}(2,\mathbb{R})'_\pm$, can be written in terms of the {\it torsion-free} Lorentzian connection $\omega^{\prime\,i}$ on the {world volume} as \,$\Omega^{\prime\,i}_\pm=\omega^{\prime\,i}\pm e^i/\ell'$. However, in line with \cite{Alvarez:2011gd} and \cite{Andrianopoli:2018ymh}, we shall eventually identify $\Omega^{\prime\,i}_{\pm}$ with $\Omega^{i}_{\pm}$ modulo additional torsion terms (which corresponds to identifying the corresponding ${\rm SL}(2,\mathbb{R})$ groups). Also, as explained in \cite{Andrianopoli:2018ymh}, the theory obtained at the boundary {of the (target space) AdS$_4$ supergravity} can be related to the model discussed in \cite{Alvarez:2011gd} by considering an Ansatz,  as in (\ref{Ansatzzanelli}), in which the gravitini are expressed in terms of the local frame and spinor fields $\chi_A=(\chi_{a_1}\,,\,\chi_{a_2})$. These are world-volume spinors in the ${\bf \left(\frac{1}{2},\,0\right)}\oplus {\bf \left(0,\,\frac{1}{2}\right)}$ representation of ${\rm SL}(2,\mathbb{R})'_+\times {\rm SL}(2,\mathbb{R})'_-$%
\footnote{{In line with the previous discussion, the $\gamma_i$ matrices in  (\ref{Ansatzzanelli})} should be thought of as intertwining matrices between target space spinor indices and world-volume spinor indices. This makes sense in light of the identification between ${\rm SL}(2,\mathbb{R})_\pm$ and ${\rm SL}(2,\mathbb{R})'_\pm$ mentioned above.}.
For the sake of notational simplicity, we use for the two sets of spinor bases $\chi_{a_1},\,\chi_{a_2}$ on which the generators of the corresponding ${\rm SL}(2,\mathbb{R})'$ groups act, to be represented by the same matrices, $i\,\gamma^i/2$. Eventually, along the lines of the discussion in Subsect.\ \ref{parity}, we shall relax this condition in order to study the parity symmetry of the theory when $p=q$.

\subsection{NYW Scale Invariance}
As shown in \cite{Alvarez:2011gd,Andrianopoli:2018ymh}, the above construction leads to the description of a propagating charged fermion satisfying a Dirac equation. Implicit in the Ansatz (\ref{Ansatzzanelli}) is the local scale invariance under the so-called Nieh-Yan-Weyl (NYW) symmetry \cite{Nieh:1981xk,Chandia:1997hu,Hughes:2012vg,Parrikar:2014usa}
\begin{align}\label{NWsymm}
e^i\to\lambda(x)\,e^i\,, \quad\;\; \chi_A\to\frac{1}{\lambda(x)}\,\chi_A\;, \qquad \lambda\neq 0\,,
\end{align}
which leaves the gravitino, and the whole theory, invariant. It is precisely the breaking of this conformal invariance that, in the framework of \cite{Guevara:2016rbl,Andrianopoli:2018ymh}, turns an originally topological Chern-Simons theory into a system with a propagating spin-1/2 field.

The identification of ${\rm SL}(2,\mathbb{R})_{\pm}$ with ${\rm SL}(2,\mathbb{R})'_{\pm}$ can be established by identifying the index $i$ of $e^i$ with the same index of $E^i$ and defines the action of the covariant derivatives $\mathcal{D}[\Omega_\pm]$ on $e^i$ as well. In particular, one can write the following general expressions for the torsion with respect to $\Omega_\pm$ as
\begin{align}\label{integrability}
T^i_{\pm}=\mathcal{D}[\Omega_\pm]e^i=\beta_\pm e^i+\tau_\pm\epsilon^{ijk} e_j\wedge e_k\,,
\end{align}
where $\beta_\pm$ and $\tau_\pm$ are 1- and 0-forms, respectively. Under the NYW symmetry transformation (\ref{NWsymm}), the above expressions retain their form provided $\beta_\pm$ and $\tau_\pm$ change according to
\begin{equation}\label{NYWbeta}
\beta_\pm\rightarrow \beta_\pm +\frac{d\lambda}{\lambda}\,,
\qquad
\tau_\pm\rightarrow \frac{1}{\lambda}\,\tau_\pm \,,
\end{equation}
that is, $\beta_\pm$ transform as a connection under local scale transformations.

Implementing the Ansaz (\ref{Ansatzzanelli}) in the ${\rm OSp}(p|2)_+\times {\rm OSp}(q|2)_-$ structure equations (\ref{AT-likecurv}) for the bosonic curvatures, yields
\begin{equation}\label{bosonicAVZ}
\begin{split}
R^i_\pm&=\pm \frac{1}{{\ell}}\,\overbar{\chi}_\pm \chi_\pm\epsilon^{ijk}\,e_j \wedge e_k\,,
\\[\jot]
\mathcal{D}[\Omega_\pm] E^i&=\mp \frac{1}{{\ell}}\,\epsilon^{ijk}\,E_j \wedge E_k+\frac{1}{2}\,(\overbar{\chi}_+\chi_++\overbar{\chi}_-\chi_-)\,\epsilon^{ijk}\,e_j \wedge e_k\,,
\\[\jot]
\mathcal{F}^{a_1 b_1}&=-\frac{i}{{\ell}}\,\left(\overbar\chi^{a_1}\gamma^i\chi^{b_1}\right)\,\epsilon_{ijk} e^j\wedge e^k\,,\qquad
\mathcal{F}^{a_2 b_2}=\frac{i}{{\ell}}\,\left(\overbar\chi^{a_2}\gamma^i\chi^{b_2}\right)\,\epsilon_{ijk} e^j\wedge e^k\,.
\end{split}
\end{equation}
where $\chi_+:=(\chi_{a_1}),\,\chi_-:=(\chi_{a_2})$. Covariantly differentiating (\ref{integrability}) yields
\begin{equation}
\mathcal{D}[\Omega_\pm]^2e^i=-\epsilon^{ijk}\,R_{\pm\,j}\,e_k=0\,,
\end{equation}
where last equality follows from the first of eqs. (\ref{bosonicAVZ}). This in turn requires $d\beta_\pm=0$ and, for non vanishing
$\tau_\pm$,
\begin{equation} \label{Beta=dlog}
\beta_\pm=-\frac{d\tau_\pm}{\tau_\pm}=-d{\rm ln}(|\tau_\pm|)\,.
\end{equation}
\sloppy
This last relation means that $\beta_{\pm}$ can be viewed as produced by the scale transformation ${e^i \rightarrow (\tau_{\pm})^{-1} e^i}$. Consequently, in the absence of global obstructions, either $\beta_+$ or $\beta_-$ can be gauged away to zero, and correspondingly either $\tau_+$ or $\tau_-$ can be set equal to a constant, by an appropriate NYW transformation.

Next we express $E^i$ in the basis of $e^i$. Consistently with our assumptions we can write the following proportionality relation
\begin{align}\label{relatingvielbein}
E^i=f\,e^i\,,
\end{align}
where $f$ is some indeterminate function.\footnote{{In general, the 1-form $E^i$ may not be parallel to $e^i$. We briefly touch on this more general case in Section 5.}} Since we have defined $\Omega^i_\pm = \omega^i \pm E^i/\ell$, assuming (\ref{relatingvielbein}), the covariant derivatives of $e^i$ can be written as \begin{equation}\mathcal{D}[\Omega_\pm]\,e^i= \mathcal{D}[\omega]\,e^i {\mp} (f/\ell)\, \epsilon^{ijk}e_j\wedge e_k\,,\end{equation} from which one obtains
\begin{equation}
 (\beta_+ - \beta_-)\,e^i +(\tau_+ -\tau_-{+}2f/\ell)\,\epsilon^{ijk}e_j \wedge e_k =0 .\label{differenceseq}
 \end{equation}

One can also compute the covariant derivative of $e^i$ with respect to $\omega^i$, which has the following general form

\begin{equation}\label{Dome}
 \mathcal{D}[\omega]e^i=\beta \wedge e^i+\tau\,\epsilon^{ijk} e_j\wedge e_k\,\,,
 \end{equation}
Comparing equation (\ref{Dome}) with  (\ref{differenceseq}), one can find
\begin{equation}\label{betatau}
\beta_+=\,\beta_-=\,\beta\,, \qquad \tau_+ +\, \frac{f}{\ell}\,=\,\tau_- -\, \frac{f}{\ell}\,=\,\tau\,.
\end{equation}
Note that in the absence of global obstructions the 1-form $\beta$ can be disposed of through a NYW transformation (\ref{NYWbeta}) \cite{Alvarez:2011gd}\footnote{Clearly, in the presence of global obstructions, this can only be done locally {in an open neighborhood of every spacetime point}.}. This would leave the theory only {invariant} under global (rigid) NYW transformations, which can in turn be used to fix the value of either $\tau_+$ or $\tau_-$ at will since, for $\beta=0$, $\tau_\pm$ are constants.

In order to find $f(x)$, we {use (\ref{relatingvielbein}) in the second expression of (\ref{bosonicAVZ}). Comparing this} with (\ref{integrability}) leads to the following conditions for $f(x)$:
\begin{align}
df+\beta \,f&=0\,,\label{eqf0}
\\
f\,\tau&=\frac{1}{2}\,(\overbar{\chi}_+\chi_++\overbar{\chi}_-\chi_-)\,,\label{eqf}
\end{align}
where $\tau$ was introduced in (\ref{Dome}).

{For $\beta=0$, eq.\ (\ref{eqf0})} is satisfied by {$f=\alpha_\pm \,\tau_\pm$, where $\alpha_\pm$ are dimensionful {constants}. An additional constraint comes} from the Bianchi identities for $R^i_\pm$, obtained from the first of eqs. (\ref{bosonicAVZ}),
\begin{equation}
\mathcal{D}[\Omega_\pm]R^i_\pm=0\;\;\,\Rightarrow\;\;\,
d(\overbar{\chi}_\pm \chi_\pm)= -2\, \beta\,\overbar{\chi}_\pm \chi_\pm\,,\label{dchichi}
\end{equation}
{Under a NYW transformation, one can always set $\beta=0$ locally. Then, the second expression in} (\ref{dchichi}) implies that, in a local patch, $\overbar{\chi}_\pm \chi_\pm$ are constants, consistently with the results of \cite{Andrianopoli:2018ymh}.
In general, for non-vanishing $\beta$, the last of eqs. (\ref{dchichi}) implies that:
\begin{equation}
\overbar{\chi}_+ \chi_+=k\,\overbar{\chi}_- \chi_-\;,
\end{equation}
with $k=$ constant.\par\smallskip
Let us now turn to the discussion of the fermionic sector of the model. {In general,} we can keep the local NYW symmetry of the theory manifest by including its connection $\beta$ in the definition of the covariant derivative and defining
\begin{equation}
\hat{\mathcal{D}}=\mathcal{D}+ w\,\beta\,,
\end{equation}
where {$w$} is the NYW weight of the field ($-1$ for $e^i$ and $+1$ for $\chi_\pm$). {Thus,} the NYW-covariant derivatives on $\chi_\pm$ {is}
\begin{equation}
\hat{\mathcal{D}}[\Omega_\pm,\,A_\pm]\,\chi_\pm=\mathcal{D}[\Omega_\pm,\,A_\pm]\,\chi_\pm+ \beta\,\chi_\pm\,,
\end{equation}
where
\begin{equation}
\mathcal{D}[\Omega_+,\,A_+]\,\chi^{a_1}_+ :=d\chi^{a_1}+\frac{i}{2}\,\Omega^i_+\,\gamma_i\chi^{a_1}+
A^{a_1b_1}\chi_{b_1}\,,
 \end{equation}
and similarly for $\mathcal{D}[\Omega_-,\,A_-]\chi_-$. {From eqs.\ \eqref{eq:DOmAPsi}} one finds
\begin{equation}\label{Dirac0}
\gamma_{[i}\hat{\mathcal{D}}_{j]}[\Omega_\pm,\,A_\pm]\,\chi_\pm=\tau_\pm \,\epsilon_{ijk}\,\gamma^k\chi_\pm\,.
\end{equation}
Contracting both sides on the left by $\gamma^{ij}$ we end up with the following Dirac equations:
\begin{equation}\label{Diracs}
 \slashed{\mathcal{D}}[\Omega_\pm,\,A_\pm]\,\chi_\pm=-3\,i\,\tau_\pm\,\chi_\pm\,,
\end{equation}
 while contracting both sides of (\ref{Dirac0}) to the left by $\gamma^i$ and using (\ref{Diracs}) one finds
\begin{equation}
 \hat{\mathcal{D}}_{i}[\Omega_\pm,\,A_\pm]\,\chi_\pm=-i\,\tau_\pm\,\gamma_i\chi_\pm\,.
 \end{equation}

Equations (\ref{bosonicAVZ}) and (\ref{Dirac0}) can all be derived from {an action for two Dirac fields, $\chi_\pm$, minimally coupled to two independent sets of CS connection fields, $\Omega_\pm,\,A_\pm$,}
\begin{equation*}
\begin{split}
S=&\int\bigg[\frac{1}{2}\left(\Omega_{+ i}\wedge d\Omega_{+}^{i}-\frac{1}{3}\epsilon^{ijk}\Omega_{+ i}\wedge \Omega_{+ j}\wedge \Omega_{+ k}\right)-\frac{1}{2}\left(\Omega_{- i}\wedge d\Omega_{-}^{i}-\frac{1}{3}\epsilon^{ijk}\Omega_{- i}\wedge \Omega_{- j}\wedge \Omega_{- k}\right)+\\
&+\left(A^{a_1b_1}\wedge dA_{b_1a_1}+\frac{2}{3}A^{a_1b_1}\wedge A^{b_1c_1}\wedge A^{c_1a_1}\right)-\left(A^{a_2b_2}\wedge dA_{b_2a_2}+\frac{2}{3}A^{a_2b_2}\wedge A^{b_2c_2}\wedge A^{c_2a_2}\right)- \\
&-\frac{2i}{\ell}\epsilon^{ijk}\overbar\chi^{a_1}\left\{\gamma_k \hat {\mathcal{D}}[\Omega_+,A^{a_1b_1}]\chi_{a_1}+i\tau_+\chi_{a_1}e_k\right\}\wedge e_i\wedge e_j-\\
&-\frac{2i}{\ell}\epsilon^{ijk}\overbar\chi^{a_2}\left\{\gamma_k \hat {\mathcal{D}}[\Omega_-,A^{a_2b_2}]\chi_{a_2}+i\tau_-\chi_{a_2}e_k\right\}\wedge e_i\wedge e_j -\frac{1}{2}\,d(\Omega_+{}_k\wedge \Omega^k_-)\bigg]\,.
\end{split}
\end{equation*}
{Note that the above action can be obtained from (\ref{Lagrangianpm}) using the Ansatz (\ref{Ansatzzanelli}).}

\subsection{{Fixing the NYW Scaling}}\label{FNYW}
As already mentioned, {the NYW symmetry can be used to set $\beta=0$ locally on any open neighborhood of the world volume. Globally,} this requires integrability of (\ref{Beta=dlog}), which imposes a nontrivial condition on the topology of spacetime. Once the scale invariance has been used to set $\beta(x)=0$, one can then use the remaining global NYW symmetry to fix either $\tau_+$ or $\tau_-$ (which are constants) to some chosen value.
It is useful to write the field equations in terms of the torsion-free Lorentz connection
$\omega^{\prime i}$
\begin{equation}
\omega^{\prime i}={\Omega_+^i}+\tau_+\,e^i={\Omega_-^i}+\tau_-\,e^i\,.
\end{equation}
The reader can easily verify that the Dirac equations in the two sectors can be recast in the form
 \begin{equation}\label{Diracs2}
 \slashed{\mathcal{D}}[\omega^{\prime },\,A_\pm]\chi_\pm=-\frac{3}{2}\,i\,\tau_\pm\,\chi_\pm\,,
 \end{equation}
where, as in the models discussed in \cite{Alvarez:2011gd, Andrianopoli:2018ymh}, the mass of the spinor fields are fixed in terms of the torsion
\begin{equation}
    \label{diracmass}
m_\pm= \frac32\, \tau_\pm\,.
\end{equation}
The Riemann tensor associated with $\omega^{\prime }$, using eq.s (\ref{bosonicAVZ}), (\ref{betatau}), (\ref{eqf0}) and (\ref{eqf}),  reads
 \begin{equation}\label{R'}
R^i[\omega^{\prime }]=\frac{1}{2}\left(\frac{f^2}{\ell^2}+\tau^2+\frac{\eta_{AB}
 \overbar{\chi}^A\chi^B}{\ell}\right)\,\epsilon^{ijk} e_j\wedge e_k\,.
 \end{equation}

For $\beta=0$, the coefficient of $\epsilon^{ijk} e_j\wedge e_k$ {in (\ref{R'}) is a constant that} defines an effective cosmological constant, which also receives a contribution from the fermion condensate (recall that $f$ and $\tau$ are related by eq.\ (\ref{eqf})). The coefficient in front of this contribution  depends on the choice of spin connection on the world volume. In particular, the residual global NYW symmetry can be used to identify the AdS$_3$ radius of the world volume, $\ell'$, with the one on the target space, $\ell$. This still allows for several choices of world-volume spin connection, which can be labeled by a real parameter $\lambda$. Identifying the gauge connection $\Omega^i_{(\lambda)}\equiv \omega^i+ \frac{\lambda}{\ell}\,E^i$ with the tangent space connection $\Omega^{\prime i}_{(\lambda)}\equiv \omega^{\prime i}+ \frac{\lambda}{\ell}\,e^i$, yields
\begin{equation}
\tau=\frac{\lambda}{\ell}(f-1) \, ,
\end{equation}
which combined with (\ref{betatau}) gives
\begin{equation}
    \tau_\pm=\frac{1}{\ell}\left[ \lambda(f-1)\mp f\right]\,.\label{taupmlambda}
\end{equation}


In this case eq.\ (\ref{eqf}) implies
\begin{equation}
\lambda\,f\,{(f-1)}=\frac{{\ell}}{2}\,(\overbar{\chi}_+\chi_++\overbar{\chi}_-\chi_-)\,.\label{ffm1p3}
\end{equation}
Under reflection $\lambda$ changes sign, so that parity invariance requires $\lambda=0$, which in turn implies $\overbar{\chi}_+\chi_+=-\overbar{\chi}_-\chi_-$. This case is alternatively described by the limit $\ell \to\infty$ of vanishing cosmological constant. The cases $\lambda=\pm 1$ correspond to the choices $\Omega^{\prime\,i}_\pm=\Omega^{i}_\pm$. In particular, for $\lambda=-1$, this identification includes the one assumed in \cite{Andrianopoli:2018ymh}, where the gauge super-group was defined by $p=0$ and $q=2$, and $\tau_-=1/{\ell}$, $\tau_+=-(2f-1)/{\ell}$.

Note that the left-hand-side of (\ref{ffm1p3}) also vanishes for $f=0$ and $f=1$. The first case can be excluded on physical grounds since it would imply $E^i=0$, which is singular. The second possibility, $f=1$, implies $\overbar{\chi}_+\chi_+=-\overbar{\chi}_-\chi_-$, which, as pointed out above, is the necessary condition for parity invariance.
Moreover one can verify that the absolute value of the left hand side of eq.\ (\ref{ffm1p3}) has a minimum for $f=1/2$. In this case, as it follows from eq.\ (\ref{taupmlambda}), choosing $\lambda$ to be $+1$ or $-1$, implies that either $\chi_-$ or $\chi_+$ are massless (i.e. $\tau_-=0$ or $\tau_+=0$, respectively).
In the next section we shall elaborate on these conditions in relation to the application of our construction to the effective description of {graphene-like systems}.

\section{Interpretation in terms of {graphene-like} 2D materials}\label{4}
In the spirit of \cite{Alvarez:2011gd} we shall discuss an application of our construction to the effective long wavelength description of the electronic properties of {graphene-like systems}. From this perspective, the spin $1/2$ fields $\chi_A$, which satisfy the Dirac equations (\ref{Diracs}), describe the electron wave-functions. Let us recall few facts about the electronic structure of {these materials}.

A graphene sheet is a two-dimensional system of carbon atoms arranged in a honeycomb lattice \cite{novoselov66,novoselov2005twodimato} (for review on the subject, see also \cite{neto2009electronic,katsnelson2007graphene,Vozmediano:2010zz,Cortijo:2006xs}). From the perspective of high energy physics, graphene provides a real framework to study Dirac pseudoparticles at sub-light speed regime \cite{neto2009electronic,novoselov2005twodimgas,Zhang:2005zz,Gusynin:2006ym,Iorio:2013ifa}, and many high-energy physics effects can be explored in a solid state system \cite{katsnelson2007graphene,Cortijo:2006xs,geim2007rise,Boada:2010sh,Iorio:2010pv,Vozmediano:2010zz,Gallerati:2018dgm,Iorio:2019czs}.

The Dirac spinorial formulation emerges from the peculiar honeycomb structure, where a unit cell is made of two adjacent atoms belonging to inequivalent sublattices, labelled A and B, respectively. This means that we find two inequivalent sites per unit cell, the distinction not referring to different kinds of atoms -- they are all carbon atoms -- but to their topological inequivalence. The single-electron wave function is then conveniently described as a two-component Dirac spinor $\zeta$ which, in a basis where the  gamma-matrices $\underline{\gamma}^i$ have the form
\begin{equation}
    \underline{\gamma}^0=-\sigma^3\,,\qquad \underline{\gamma}^1=-i\,\sigma^2\,,\qquad \underline{\gamma}^2=i\,\sigma^1\,,
\end{equation}
can be written as
\begin{equation}
  \zeta=\left(\begin{array}{c}
    \sqrt{n_{A}}\, e^{i\alpha_A} \\
    \sqrt{n_{B}}\, e^{i\alpha_B}
\end{array}\right)\,. \label{zetaAB}
\end{equation}
Here $n_{A}$, $n_{B}$ are the probability densities for the electron in the $\pi$-orbitals, referred to the A and B sublattices, respectively, and $\alpha_A,\alpha_B$ the corresponding wave-function phases. In terms of $\zeta$ one can define the following two quantities:
\begin{equation} \label{n-dn}
n\equiv n_A+n_B=\zeta^\dagger \zeta\,,\qquad
\Delta n\equiv n_B-n_A= \bar{\zeta}\zeta\,,
\end{equation}
where $n$ is the total electron probability density while $\Delta n$
is the asymmetry in the probability density between the two sublattices. This description is robust under changes of the lattice preserving the topological structure.

The Dirac physics is realized for low-lying energy pseudoparticle excitations: for energy ranges where the electron wavelength is much larger than the lattice length, the charge carriers see the graphene sheet as a continuum 2+1 dimensional spacetime. Moreover, quasiparticles with large wavelength are sensitive to sheet curvature effects, calling for a quantum Dirac field formulation in curved spacetime \cite{Boada:2010sh,Gallerati:2018dgm}.

{Let us recall that isolated pristine graphene features massless Dirac equations for the pseudoparticles at the Dirac points. However, mass terms can be induced in several ways, for instance by switching on suitable local magnetic fluxes (see Appendix \ref{App_B}). Moreover, other graphene-like 2D materials exist where parity symmetry between the A and B sites is absent, and a mass gap is present, due to the different kind of atoms in the honeycomb lattice. This is the case, for instance, of the boron nitride, where effective parity-violating mass terms emerge \cite{Semenoff:1984dq}.\footnote{{A possibility for producing a parity violating mass gap in a graphene monolayer is to deposit it on a suitable substrate, for instance of boron nitride \cite{Giovannetti2007substrate} or silicon carbide \cite{zhou2007substrate}, inducing in this way local on-site potentials spoiling the original parity invariance between A and B sites.}}}\par\smallskip

The relation between the spin-$1/2$ fields of our model and $\zeta$ can be stated as follows:
\begin{equation}
\chi=\sqrt{\frac{\ell}{2}}\,U\,\zeta\,, \label{chizeta}
\end{equation}
where the dimensionful constant $\sqrt{\frac{\ell}{2}}$ is needed in order for $\zeta$ to have the correct dimension of $1/{\rm (length)}$ and the $2\times 2$ matrix $U$ relates the spinor basis used for $\chi$  {(see Appendix \ref{App_A})} to the one defined above for $\zeta$: \,$U^\dagger \gamma^i U=\underline{\gamma}^i$. The matrix $U$ is readily found to be
\begin{equation}
    U=\frac{1}{\sqrt{2}}\left(\begin{matrix}1 & 1 \cr -i & i\end{matrix}\right)\,.
\end{equation}
We shall restrict ourselves to the case in which supersymmetry is defined by even integers $p$ and $q$, since this allows to arrange the real spinors $\chi_{\pm}$  into ${p}/2$ and $q/2$ Dirac spinors. The simplest choice would be the case $p=2,\,q=0$ or $p=0,\,q=2$, discussed in \cite{Alvarez:2011gd} and \cite{Andrianopoli:2018ymh}. The next simplest case corresponds to $p=q=2$, which will be discussed next.

An important consequence of (\ref{chizeta}), as first shown in \cite{Andrianopoli:2018ymh} and derived on general grounds in the previous section, is that the quantity \begin{equation}\label{chichinanb}
\overbar{\chi} \chi =\frac{\ell}{2}\,\bar{\zeta} \zeta=\frac{\ell}{2}\,(n_B-n_A)\,,
\end{equation}
 by virtue of the last of eqs. (\ref{dchichi}), is constant for $\beta=0$, in which case the difference $n_B-n_A$ in the probability densities is a constant index whose relevance will be further explored in future work.

Let us elaborate now on a consequence of eq.\ (\ref{chichinanb}) in light of the discussion in subsection \ref{FNYW}.
Equation (\ref{ffm1p3}) implies the following bounds:
\begin{equation}
\begin{split}
&\lambda >0\;,\quad n_A=-\frac{4\lambda}{\ell^2}\,f(f-1)+n_B\ge -\frac{4\lambda}{\ell^2}\,f(f-1)\ge \frac{\lambda}{\ell^2}\,,\\
&\lambda <0\;,\quad n_B=\frac{4\lambda}{\ell^2}\,f(f-1)+n_A\ge \frac{4\lambda}{\ell^2}\,f(f-1)\ge \frac{|\lambda|}{\ell^2}\,.
\end{split}
\end{equation}
In the two cases we find a lower bound in the probability densities of one of the two sublattices.

In our model the spinors are split into the $+$ and $-$ sectors and $\overbar{\chi} \chi$ in (\ref{chichinanb}) should be understood as
\begin{equation}\overbar{\chi} \chi\equiv\overbar{\chi}_+ \chi_++\overbar{\chi}_- \chi_- =\frac{\ell}{2}\,(n_B-n_A)\,,
\end{equation}
so that we can write
\begin{equation}
n_A=n^{(+)}_A+n^{(-)}_A\;,\quad\;
n_B=n^{(+)}_B+n^{(-)}_B\,,
\end{equation}
where $n^{(\pm)}_A,\,n^{(\pm)}_B$ are the probability densities related to the A and B sublattices in the $+$ and $-$ sectors. In the context of {graphene-like systems}, the $+$ and $-$ sectors, {in the $p=q$ case, can be interpreted} as referring to the ${\bf K},\,{\bf K}'$ valleys. This will be discussed in the next subsection. For $p\neq q$ the interpretation of the model is more obscure and we shall put forward possible interpretations in a particular case. Note that the {case $\lambda =0$, which implies $\overbar{\chi}_+\chi_+=-\overbar{\chi}_-\chi_-$, and $n_A=n_B$,} can be realized as a non-trivial relation between the probability densities in the $+$ and $-$ sectors.

\subsection{The \texorpdfstring{$p=q$}{p=q} case and the \texorpdfstring{$\bf{K}$, $\bf{K'}$}{\bf K,K'} Dirac points}
Let us now restrict to the case $p=q$, in which the parity symmetry discussed in Sect.\ \ref{parity} emerges in the model. Since $E^i=f\,e^i$, the action of the $\mathcal{O}_Y$-parity  on $E^i$ naturally extends to $e^i$:
\begin{equation}
    e^i\,\rightarrow\, \tilde{e}^i=\mathcal{O}_Y{}^i{}_j\,e^j\,,\label{parity15}
\end{equation}
provided $f$ is invariant:  $\tilde{f}=f$. Consistency of eqs. (\ref{parity1}), (\ref{parity2}) and (\ref{parity15}) with the Ansatz (\ref{Ansatzzanelli}) implies the following transformation rule for the spin-1/2 fields
\begin{align}
\chi_\pm\,\rightarrow\, \tilde{\chi}_\pm=-\sigma^1\chi_\mp\,.
\label{parity3}
\end{align}
One can  verify for instance that $e_i\,\gamma^i\chi_{\pm}=\tilde{e}_i\,\sigma^1\gamma^i\tilde{\chi}_{\mp}$, consistently with the transformation property of the $\psi_\pm$. Invariance under reflections of the expression (\ref{integrability}) for torsion implies
\begin{equation}
\beta\,\rightarrow\,\tilde \beta  ={\beta}\,, \qquad \tau_\pm\, \rightarrow\, \tilde\tau_\pm= -\tau_\mp\;.
\end{equation}
A specific world-volume background, characterized by certain torsion components, is parity invariant provided:
\begin{equation}
\tilde{\tau}_\pm={\tau}_\pm\;\;\Rightarrow\;\;\,\tau_+=-\tau_- \;\;.
\label{Pinvariancetorsion}
\end{equation}
It is also straightforward to verify that, {under reflections},
\begin{equation}
\eta_{AB}\,\overbar{\chi}^A\chi^B = \overbar{\chi}_+\chi_+-
\overbar{\chi}_-\chi_-
\quad\;\text{and}\quad\; \overbar{\chi}\chi = \overbar{\chi}_+\chi_++
\overbar{\chi}_-\chi_-\; ,
\end{equation}
{are a scalar and a pseudo-scalar, respectively. Moreover,} the field equations are invariant while the Lagrangian density is, as expected, odd. Equations (\ref{parity1}), (\ref{parity2}) and (\ref{parity15}) implement the reflection symmetry over the Y-axis on the tangent space to the world volume. In particular the two Dirac equations (\ref{Diracs2}) are mapped into one another. An analogous discussion applies to the reflection over the X-axis in the tangent space to the world volume (see last paragraph of Sect.\ \ref{parity}). In this case the transformation properties of the spinors are: ${\chi_\pm\rightarrow \tilde{\chi}_\pm=-\sigma^3\chi_\mp.}$

The $\pm$ sectors, which are related by a reflection symmetry in one spatial axis, can be naturally associated with the ${\bf K},\,{\bf K}'$ valleys of graphene.
To motivate this, we recall that ${\bf K},\,{\bf K}'$ are the two inequivalent points in the first Brillouin zone (FBZ) of the reciprocal lattice, which also has a honeycomb geometry. We describe the elementary hexagons related to the honeycomb lattice and to the FBZ as in the figure below.
%
\begin{figure}[H]
\centering
\includegraphics[width=0.9\textwidth]{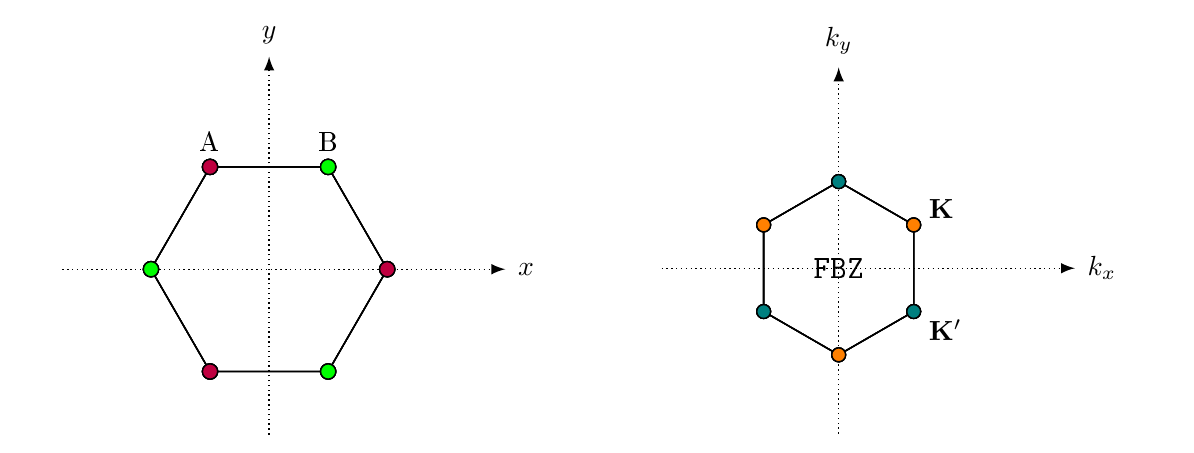}
\end{figure}
\sloppy
The relativistic behavior of the charge carriers can be inferred, in the momentum space, from the linear dispersion relation between the energy and the quasi-momentum at the vertices of the first Brillouin zone. Close to the Fermi energy, the electrons in graphene have linear energy bands, {like relativistic massless particles and} the three dimensional plot of these two dimensional bands produce the so-called Dirac cones. For every momentum lying within the Brillouin zone, the Hamiltonian has two eigenvalues with opposite signs: the positive (negative) eigenvalue corresponds to the conduction (valence) band of graphene. The conduction and valence bands touch each other at the conical apices, located at the corners of the hexagonal FBZ. The latter are then split into two equivalence classes (referred to as “valleys") and, as a result, electrons in graphene possess an additional pseudo-spin number, the \emph{valley}. {These properties are shared with other 2D graphene-like materials, which further allow the inclusion of effective mass terms.}

Consistently with our description of the honeycomb lattices (see figure above), the reflection with respect to the Y-axis exchanges the A and B sites. The points ${\bf K}$ and ${\bf K}'$ are mapped into each other if the reflection is combined with a time-reversal transformation, so that the resulting effect on a momentum vector is ${k_x\rightarrow k_x\,,\;k_y\rightarrow -k_y}$\,. {As mentioned above, this symmetry, which is present in pure graphene, is absent in 2D materials with inequivalent A and B sites. This feature implies the presence, for such materials, of a parity-violating Semenoff mass term in the effective Dirac equation.}

In the absence of curvature, it is known that the Dirac equations in momentum space in the two valleys, in our conventions, read (setting $\hbar=v_\textsc{f}=1$) \cite{neto2009electronic}:
\begin{align}
\begin{split}
&{\bf K}\: :\qquad E_{{\bf q}}
\;\chi_{{}_{\bf K}}({\bf q})=\left(\alpha^1\,q^1+\alpha^2\,q^2+ m_{{}_{\bf K}}\,\gamma^0 \right)\chi_{{}_{\bf K}}({\bf q})\;,
\\
&{\bf K}':\qquad E_{{\bf q}}
\;\chi_{{}_{{\bf K}'}}({\bf q})=\left(\alpha^1\,q^1-\alpha^2\,q^2+ m_{{}_{{\bf K}'}}\,\gamma^0 \right)\chi_{{}_{{\bf K}'}}({\bf q})\;,
\label{KKp}
\end{split}
\end{align}
where $\alpha^\ell\equiv \gamma^0\,\gamma^\ell$ ($\ell=1,2$) and the two equations are computed in the two-momenta ${\bf K}+{\bf q}$ and ${\bf K}'+{\bf q}$, with ${|{\bf q}|\ll |{\bf K}|,  |{\bf K}'|}$. The  Hamiltonian matrices on the right hand sides of equations (\ref{KKp}) have eivengalues $+|E_{{\bf q}}|$ and $-|E_{{\bf q}}|$. The above two equations, in configuration space, read:
\begin{align}
\begin{split}
&{\bf K}\: :\qquad i\,\partial_t\chi_{{}_{\bf K}}({  x})=\,\left(-i\alpha^1\,\partial_x-i\alpha^2\,\partial_y+ m_{{}_{\bf K}}\,\gamma^0 \right)\chi_{{}_{\bf K}}({x})\;,
\\
&{\bf K}':\qquad i\,\partial_t
\chi_{{}_{{\bf K}'}}(x)=\left(-i\alpha^1\,\partial_x+i\alpha^2\,\partial_y+ m_{{}_{{\bf K}'}}\,\gamma^0 \right)\chi_{{}_{{\bf K}'}}(x)\;,
\label{KKpx}
\end{split}
\end{align}
By general covariance, the generalization of equations (\ref{KKpx}) to a curved background, in the presence of minimal couplings to a gauge potential, is obtained by replacing partial derivatives by covariant ones:
\begin{align}
\begin{split}
&{\bf K}\: :\qquad i\,\mathcal{D}_t[\omega',\,A_{{}_{\bf K}}]\,\chi_{{}_{\bf K}}(x)=\,\left(-i\alpha^1\,\mathcal{D}_x[\omega',\,A_{{}_{\bf K}}]-i\alpha^2\,\mathcal{D}_y[\omega',\,A_{{}_{\bf K}}]+ m_{{}_{\bf K}}\,\gamma^0 \right)\chi_{{}_{\bf K}}(x)\;,
\\
&{\bf K}':\qquad i\,\mathcal{D}_t[\omega',\,A_{{}_{{\bf K}'}}]
\,\chi_{{}_{{\bf K}'}}(x)=\,\left(-i\alpha^1\,\mathcal{D}_x[\omega',\,A_{{}_{{\bf K}'}}]+i\alpha^2\,\mathcal{D}_y[\omega',\,A_{{}_{{\bf K}'}}]+ m_{{}_{{\bf K}'}}\,\gamma^0 \right)\chi_{{}_{{\bf K}'}}(x)\;,
\label{KKpxD}
\end{split}
\end{align}
where $A_{{}_{{\bf K}}}(x)$ and $A_{{}_{{\bf K}'}}(x)$ denote the configuration-space representation of the  gauge fields about the two Dirac points.
By comparing (\ref{KKpxD}) with eqs. (\ref{Diracs2}),
we can consistently identify the spinor field $\chi_\pm(x)$ with $\chi_{{}_{\bf K}}(x),\,\chi_{{}_{{\bf K}'}}(x)$, up to an overall normalization, as follows
\begin{equation}
\chi_{{}_{\bf K}}(x^\mu)= \chi_+(x^\mu)\;,\qquad
\chi_{{}_{{\bf K}'}}(x^0,\,x^1,\,x^2)= \sigma^1\chi_-(-x^0,\,x^1,\,x^2)\;,
\end{equation}
provided we also identify:
\begin{equation}
A_{{}_{{\bf K}}}=A_+\,,\quad\;\;
A_{{}_{{\bf K}'}}=A_-\,,
\end{equation}
and the mass gaps at the two valleys with the mass parameters $m_\pm$ of $\chi_\pm$, see eq.\ (\ref{diracmass}),
\begin{equation}
m_{{}_{\bf K}}=m_+=\frac{3}{2}\,\tau_+\;,
\qquad
m_{{}_{{\bf K}'}}=m_-=\frac{3}{2}\,\tau_-\;.
\label{midentification}
\end{equation}

This motivates the identification of the $\pm$ sectors in our model with the two valleys, and the corresponding mass gaps with the torsion parameters of our model. \par
Note that applying spatial reflection with respect to the X-axis maps eqs.\ (\ref{KKp}) into each other, provided $m_{{}_{{\bf K}'}}=-m_{{}_{\bf K}}$\,. This implies that $m_{{}_{{\bf K}'}}+m_{{}_{\bf K}}$ is parity-odd, while $m_{{}_{{\bf K}'}}-m_{{}_{\bf K}}$ is parity-even.

\subsubsection{{Microscopic interpretation}}
Mass terms can be included in {graphene-like systems} by generalizing the tight binding microscopic model (see Appendix \ref{App_B}) and opening mass gaps at the Dirac points. The generation of a gap at Dirac points was first discussed in 1984 by Semenoff, introducing a mass term through an on-site  deformation $\pm M$ breaking sublattices equivalence \cite{Semenoff:1984dq}.

Another model was proposed by Haldane, with the introduction of a periodic local magnetic flux density with zero net flux over the honeycomb hexagon \cite{Haldane:1988zza}. The corresponding physical system is represented introducing in the microscopic Hamiltonian second-neighbor hopping terms with a phase factor {$e^{\pm i\varphi}$,} the phase sign according to the ``chirality'' of the electron path, i.e.\ depending on whether the hopping is clockwise or anticlockwise w.r.t.\ the hexagonal cell. The phase $\varphi$ (Aharonov–Bohm phase) induced by the local fluxes can be taken as a parameter of the model. As discussed in Appendix \ref{App_B}, the degeneracy of the bands at the Dirac points is lifted either by non-zero $M$ or non-zero  Aharanov-Bohm phase contribution, and the fermion masses in the two inequivalent valleys turn out to be
\begin{equation}
m_{{}_{\bf K}}=M- 3\sqrt 3\, t_2 \sin{\varphi}\;,\qquad
m_{{}_{{\bf K}'}}=M+ 3\sqrt 3\, t_2 \sin{\varphi}\;,
\label{sh}
\end{equation}
where the coefficient $t_2$ is the hopping amplitude {between next-to-nearest neighbors \cite{Haldane:1988zza}} (see Appendix \ref{App_B}).

Using eq.\ (\ref{midentification}) the physical quantities expressed by the Semenoff local potential term $M$ and Haldane contribution $3\sqrt{3}\,t_2\sin{\varphi}$ {can be related} to the fermion masses $m_\pm$ of our macroscopic model. To see this, let us use eqs.\ (\ref{betatau}) to write:
\begin{equation}
\tau_\pm \equiv \frac 12\,(\tau_++\tau_-)\,\pm\,\frac 12\,(\tau_+-\tau_-)\,=\,\tau \mp\,2\,\frac {f}{\ell}\,.\label{mpmtpm}
\end{equation}

As discussed above (see in particular eq. (\ref{Pinvariancetorsion})), the  first term is parity odd, while the second is parity even. From eqs. (\ref{midentification}), (\ref{sh}) and (\ref{mpmtpm}), the following identification can be made:
\begin{equation}
M=\frac 32\tau\;,\quad\;\;
\sqrt 3\,t_2 \sin(\varphi)=\frac{ f}{\ell}\;.
\label{identification}
\end{equation}
This is consistent with the property of the Semenoff and Haldane mass terms to be parity odd and even, respectively. Indeed, parity is broken by a Semenoff-type contribution, that could be generated by an asymmetry between the A and B sites. Hence, the action of parity amounts to an interchange of $n_A$ with $n_B$. The above identification is therefore consistent with the fact that, in light of eq.\ (\ref{eqf}), $\tau$ is proportional to $\overbar{\chi}\chi$ which is in turn proportional to $n_A-n_B$.

As for the Haldane-type mass contribution, the identification (\ref{identification}) suggests a relation between the ratio $f/\ell$ and the Berry phase parameters. Its meaning deserves a separate investigation which we leave for a future work.


\subsection{A different model for graphene}
Let us now briefly discuss a different application of our model to {graphene-like materials}, which makes contact with the analysis in \cite{Iorio:2018agc}. In the latter the authors generalize the original model of \cite{Alvarez:2011gd} by considering a superalgebra of the form $A(1,1)={\rm SU}(2|1,1)$ whose bosonic subgroup is ${\rm SU}(1,1)\times {\rm SU}(2)$, with respect to which the supersymmetry generators have the following transfomation property:
\begin{equation}
    Q_{IA\,\alpha}\,\in\,\,{\bf \left(\frac{1}{2},\frac{1}{2}\right)}\oplus{\bf \left(\frac{1}{2},\frac{1}{2}\right)}\,,
\end{equation}
where $I=1,2$ is a ``flavour'' index labelling the two irreducible representations in the above direct sum, $A$ and $\alpha$ run over the doublet representations of $ {\rm SU}(2)$ and ${\rm SU}(1,1)$, respectively. Applying the AVZ Ansatz to the spinor 1-forms $\psi^{IA\,\alpha}$ dual to $ Q_{IA\,\alpha}$, one ends up with four real 2-spinors $\chi^{IA}=(\chi^{IA\,\alpha})$ which can be grouped into two Dirac ones $\chi^A\equiv \chi^{1\,A}+i\,\chi^{2\,A}$. In \cite{Iorio:2018agc} the doublet index $A$ refers to the ${\bf K}$ and ${\bf K}'$ valleys and the ${\rm SU}(2)$ internal symmetry group is naturally gauged by construction. This allows the authors to describe topological features of graphene such as \emph{grain boundaries}. We emphasize here that this theoretical construction is substantially different from the one described in the previous subsection, in that the ``valley'' pseudo-spin is not associated with two different ``$\pm$'' sectors of the AdS$_3$ Achucarro-Townsend supergravity, but with just one of the two, since no supersymmetry is assumed in the other (as it is the case for the orignal model of \cite{Alvarez:2011gd}). We also emphasize that the approach pursued here is a \emph{top-down} one.\par
Although a supergroup of the form ${\rm SU}(2|1,1)_+\times {\rm SO}(2,1)_-$ is not comprised in our class of models, the closest we can get to the construction of \cite{Iorio:2018agc} corresponds to choosing $p=4$ and $q=0$. In this case the supergroup we start from is ${\rm OSp}(4|2)_+\times {\rm SO}(2,1)_-$, whose bosonic subgroup is ${\rm SO}(4)\times {\rm SL}(2,\mathbb{R})_+\times {\rm SL}(2,\mathbb{R})_- $\,. By writing the R-symmetry group in the equivalent form ${\rm SO}(4)={\rm SU}(2)_1\times {\rm SU}(2)_2$,
the supercharges can be characterized as transforming in the following real irreducible representation
\begin{equation}
    Q_{IA\,\alpha}\,\in\,\,{\bf \left(\frac{1}{2},\frac{1}{2},\frac{1}{2}\right)}\,,
\end{equation}
where now $I=1,2$ and $A=1,2$ are the doublet indices of the two factors ${\rm SU}(2)_1$ and ${\rm SU}(2)_2$ of the R-symmetry group. Consequently, upon implementing the AVZ Ansatz,
the index structure of the resulting spinors is $\chi^{I A \alpha}$, formally the same as in the model of \cite{Iorio:2018agc} described above, with the difference that now the index $I$ is no longer a flavour one but it is acted on by the gauge group ${\rm SU}(2)_1$ which was absent in the construction of \cite{Iorio:2018agc}, while ${\rm SU}(2)_2$, acting on the $A$ index, should be identified with the R-symmetry group of \cite{Iorio:2018agc}. Therefore we expect that most of the applications of the model of \cite{Iorio:2018agc} to the study of the topological features of graphene should also hold for our ${\rm OSp}(4|2)_+\times {\rm SO}(2,1)_-$ model. The main difference is the presence in the latter of additional gauge vectors $A^{IJ}_\mu=A^{JI}_\mu$, associated with ${\rm SU}(2)_1$. It is tempting to identify these vectors with the true spin-connection, and the corresponding group ${\rm SU}(2)_1$ with the true spin of the $\pi$-electrons in the {honeycomb} layer.
From the Maruer-Cartan equations of the superalgebra, upon using the AVZ Ansatz, we derive, apart from the Dirac equations for $\chi^{I A \alpha}$, the following equations for the curvature and gauge field strengths:
\begin{align}
    R_+^i&=\frac{i}{\ell}\,{\chi}^{IA\alpha}\chi^{JB\beta}\,\epsilon_{IJ}\,\epsilon_{AB}\,\epsilon_{\alpha\beta}\,\epsilon^{ijk}\, e_j\wedge e_k\,,\nonumber\\
\mathcal{F}^{AB}&\equiv dA^{AB}-\frac{1}{2}\,\epsilon_{CD}\,A^{AC}\wedge A^{DB}= -\frac{i}{\ell}\,{\chi}^{IA\alpha}\chi^{JB\beta}\,\epsilon_{IJ}
(C\gamma^i)_{\alpha\beta}\,\epsilon_{ijk} \,e^j\wedge e^k\,,\nonumber\\
\mathcal{F}^{IJ}&\equiv dA^{IJ}-\frac{1}{2}\,\epsilon_{KL}\,A^{IK}\wedge A^{LJ}= -\frac{i}{\ell}\,{\chi}^{IA\alpha}\chi^{JB\beta}\,\epsilon_{AB}
(C\gamma^i)_{\alpha\beta}\,\epsilon_{ijk}\, e^j\wedge e^k\,,\nonumber
\end{align}
where $\mathcal{F}^{IJ}=\mathcal{F}^{JI}$ and $\mathcal{F}^{AB}=\mathcal{F}^{BA}$ are the field strengths associated with the internal symmetry groups  ${\rm SU}(2)_1$ and ${\rm SU}(2)_2$, respectively. Interpreting the former as the true spin group, the corresponding field strength would describe the true spatial curvature of the {two-dimensional} sheet.\par Note that if we choose the fermion field $\chi$ to have the following special ``factorized'' form $\chi^{IA\alpha}=v^{IA} \chi^\alpha$, where $\chi^\alpha$ are Grassmann numbers while $v^{IA}$ are real ones,  in the above equations $\mathcal{F}^{AB}=\mathcal{F}^{IJ}=0$ and the fermion field ceases to be a source for the gauge field strengths.
We refrain from studying the features of this model any further here, leaving it to a future work.

\section{Concluding remarks}
Here we have seen how the extension of unconventional supersymmetry to the ${\rm OSp}(p|2)\times {\rm OSp}(p|2)$ superalgebra can be instrumental in describing the electronic properties of {graphene-like systems} in the ${\bf K}$ and ${\bf K}'$ valleys and thus physical situations in which the symmetry between them is broken. These can be realized, for instance,  by breaking reflection or time-reversal symmetries through the Semenoff or Haldane-type mass terms, as produced by the presence of suitable substrate and magnetic fields. One of the main results of this work was to embed this effective description in an $\mathcal{N}$-extended four-dimensional supergravity.
This sets the stage of a holographic analysis which will be pursued in a future work.

There is a different application of our construction to the description of graphene, which makes contact with the work by Iorio and Pais\ \cite{Iorio:2018agc}, and which we just touched upon here.

As a final comment, we observe that a graphene sheet is "relativistic" in the sense of Fermi velocity $v_\textsc{f}$ playing the role of analogue speed of light for the charge carriers. However, in our top-down approach the speed of light, as  coming from the $D=4$ supergravity model, is naturally identified with the true speed of light, $c$.
Actually, this issue can be dealt with in different ways. We could either think of the $D=4$ supergravity as already analogue, or we could instead define a more general relation between the two frames $E^i$ and $e^i$  than the one assumed here in (\ref{relatingvielbein}), of the form $ E^i=f\,\mathpzc{M}^i_{\,\,j}\,e^j$. By choosing, for instance, $\mathpzc{M}= {\mathrm{diag}}(\alpha^2,1,1)$, we can introduce an analogue speed of light in the world volume $\hat c=\alpha \,c$.
The mathematical implications of such assumption are under investigation  by one of the authors \cite{Noris:2019sdw}.\par
We also leave to a future investigation the construction of explicit solutions, including topologically non-trivial configurations, to the equations of our model and the derivation from them of explicit phenomenological predictions.

\section*{\bf{Acknowledgements}}
We are grateful to A.\ Anabalón, F.\ Dolcini, David J.\ Fern\'andez, R.\ Olea and L.\ Ravera for useful discussions. {This work has been partially funded by Fondecyt Grants 1161311 and 1180368. The Centro de Estudios Cien\'ificos (CECs) is funded by the Chilean Government through the Centers of Excellence Base Financing Program of Conicyt.}

\appendix

\section{\texorpdfstring{$\textrm{OSp}(\mathcal{N}|4)$}{\textrm{OSp}(N|4)} algebra and conventions} \label{App_A}

In this appendix we review the ortho-symplectic superalgebra and, while doing so, we state the conventions used in this paper.\par
The whole algebra is given by the following relations
\begin{equation}\label{wholealgebra}
\begin{split}
[L_{\mathcal{A}\mathcal{B}},L_{\mathcal{C}\mathcal{D}}]&=\kappa_{\mathcal{A}\mathcal{D}}L_{\mathcal{B}\mathcal{C}}-\kappa_{\mathcal{A}\mathcal{C}}L_{\mathcal{B}\mathcal{D}}+\kappa_{\mathcal{B}\mathcal{C}}L_{\mathcal{A}\mathcal{D}}-\kappa_{\mathcal{B}\mathcal{D}}L_{\mathcal{A}\mathcal{C}}, \\[\jot]
[T_{AB},T_{CD}]&=\delta_{AD}T_{BC}-\delta_{AC}T_{BD}+\delta_{BC}T_{AD}-\delta_{BD}T_{AC}, \\[\jot]
[L_{\mathcal{A}\mathcal{B}},Q_A^\alpha]&=-\frac{1}{2}(\tilde\Gamma_{\mathcal{A}\mathcal{B}})^\alpha{}_\beta Q_A^\beta,\\[\jot]
[T_{AB},Q_C^\alpha]&=2\delta^D{}_{[A}\delta_{B]C}Q_D^\alpha, \\[\jot]
\{Q_A^\alpha,Q_B^\beta\}&=\frac{1}{2\ell}(\tilde\Gamma^{\mathcal{E}\mathcal{F}}C_5)^{\alpha\beta}\delta_{AB}L_{\mathcal{E}\mathcal{F}}-\frac{1}{\ell}C_5^{\alpha\beta}T_{AB}.
\end{split}
\end{equation}
The first two properties describe the bosonic subalgebra of the subgroup $\textrm{O}(\mathcal{N})\times \SO(2,3)$,
where
\begin{equation}
\begin{split}
\mathcal{A},\mathcal{B},\dots&=0,1,2,3,4\,;\\[1.5\jot]
\kappa_{{}_{\mathcal{A}\mathcal{B}}}&=\text{diag}(+,-,-,-,+)\,,
\end{split}
\end{equation}
and
\begin{align}
A,B,\dotsc=1,\dots,\mathcal{N}\,.
\end{align}
The other three relations extend the bosonic subalgebra to a supersymmetric one and necessarily involve the fermionic generators $Q_A^\alpha$: they are Majorana spinors in the fundamental representation of $\SO(\mathcal{N})$. Being spinors, they also have a Lorentz index $\alpha$ in the spinorial representation, which means that $\alpha,\beta, \ .\ .\ .=0,1,2,3,4$. One can show that all these relations satisfy Jacobi identities, which in turn means that this is indeed an algebra.\par
We now clarify the conventions for the Dirac matrices: if $i=0,1,2$, then
\begin{equation}
\begin{split}
\Gamma^i&=\sigma_1\otimes\gamma^i\,, \quad \gamma^0=\sigma_2\,, \quad \gamma^1=i\sigma_1\,, \quad \gamma^2=i\sigma_3\,, \\[\jot]
\Gamma^3&=i\sigma_3\otimes\textbf{1}\,,\quad
\Gamma^5=i\,\Gamma^0\Gamma^1\Gamma^2\Gamma^3=-\sigma_2\otimes\textbf{1}\,.
\end{split}
\end{equation}
with
\begin{align}
\{\Gamma^a,\Gamma^b\}=2\kappa^{ab}\textbf{1}_{4\times4}
\end{align}
where $\kappa^{ab}=\text{diag}(+,-,-,-)$ and $a,\,b,\dotsc =0,1,2,3$. \par
The $\tilde\Gamma_\mathcal{A}$ matrices appearing in \eqref{wholealgebra} are related to the $D=4$ gamma matrices $\Gamma_a$ by
\begin{align}
\tilde\Gamma_a:=i\Gamma_a\Gamma_5, \qquad \tilde\Gamma_4:=\Gamma_5.
\end{align}
in such a way that they satisfy
\begin{align}
\tilde\Gamma_{ab}=\Gamma_{ab}, \qquad \tilde\Gamma_{a4}=i\Gamma_a.
\end{align}
At last, the charge conjugation matrix $C_5$ appearing in the algebra is defined as
\begin{align}
C_5:=\tilde\Gamma_0\tilde\Gamma_4=\Gamma_0
\end{align}
with the straightforward properties
\begin{align}
C_5=C_5^{-1}=-C_5^t=-C_5^*, \qquad C_5^{-1}\tilde\Gamma_\mathcal{A}C_5=(\tilde\Gamma_\mathcal{A})^t.
\end{align}
Notice that $C_5$ behaves as the usual 4-dimensional charge conjugation when acting on $\Gamma_a$ matrices
\begin{align}
C_5^{-1}\Gamma_aC_5=-(\Gamma_a)^t.
\end{align}
In these cases we will just indicate the charge conjugation matrix as $C$. \par
Finally, the Dirac conjugate of a 4-d spinor is given by
\begin{align}
\bar\Psi=-i\Psi^\dag\tilde\Gamma_0\tilde\Gamma_4=\Psi^\dag\Gamma_0,
\end{align}
whereas a Majorana spinor satisfies the reality property
\begin{align}
\Psi=-C_5\Psi^t=\Psi^*.
\end{align}
\paragraph{Conventions in D=3\\}
We choose the mostly minus convention for the signature of the three dimensional spacetime and $\epsilon^{012}=\epsilon_{012}=1$. The Lorentz covariant derivatives are defined on vectors $E^i$ as
\begin{align}
 \mathcal{D}[\omega]E^i:=dE^i+\omega^{ij}E_j=dE^i-\epsilon^{ijk}\omega_jE_k\,,
\end{align}
where $\omega^{i}:=\frac{1}{2}\epsilon^{ijk}\omega_{jk}$, while on spinors $\psi$ as
\begin{align}
    \mathcal{D}[\omega]\psi:=d\psi+\frac{1}{4}\omega^{ij}\gamma_{ij}\psi=d\psi+\frac{i}{2}\omega^i\gamma_i\psi.
\end{align}

\section{Microscopic description for {graphene-like systems}} \label{App_B}
Pure graphene quantum states can be formulated in terms of the so-called \emph{tight-binding model}, describing electrons hopping in the (single-state per site) honeycomb lattice. In the limit of very far apart ions, the single-particle eigenstates refer to an electron affected by a single ion, resulting in a set of lattice sites with a single-level state.
Within this model, electrons can tunnel to their first neighbor atoms, with a hopping amplitude $t_1$ (for graphene one has $t_1\simeq-2.7\,\text{eV}$).
The electronic system is described by the single orbital, tight-binding Hamiltonian:
\begin{equation}
{H}_1=t_1\,\sum_{\langle i,j\rangle}\, c_i^\dagger\,c_j\;,
\label{Ham1}
\end{equation}
where the creation (annihilation) operator $c_i^\dagger=c^\dagger(\mathbf{r}_i)$ \big($c_i=c(\mathbf{r}_i)$\big) acts on particle site $\mathbf{r}_i$, and the sum $\langle i,j\rangle$ runs on nearest neighbors sites $\mathbf{r}_i$, $\mathbf{r}_j$.\par\smallskip
%
\subsection{Massive deformations}
The above massless formulation is in general robust, since it comes out at the level of non-interacting system and is protected by combination of parity and time-reversal symmetry of the framework. However, mass terms {for 2D, graphene-like systems} can be obtained from generalization of a tight binding microscopic model, opening mass gaps at the Dirac points.
This gap generation was first discussed by Semenoff, introducing a mass term through an on-site staggered potential $\pm M$ spoiling sublattices equivalence \cite{Semenoff:1984dq} and breaking parity symmetry of the theory.
Another model was proposed by Haldane, including local magnetic fields over the honeycomb hexagon, breaking time-reversal symmetry of the model.

\paragraph{Haldane model.}
The formulation of the Haldane model \cite{Haldane:1988zza} was motivated by the realization of a quantum anomalous Hall effect (topological, quantized insulating phase), in the absence of Landau level structure. This can be achieved by the introduction of periodic local magnetic flux densities, with zero net flux over the cell.
The physical system is represented introducing in the microscopic Hamiltonian second-neighbor hopping terms with unimodular phase factor, the phase sign depending on the ``chirality'' of the electron path (according to whether the hopping is clockwise or anticlockwise w.r.t.\ the cell).
The microscopic Hamiltonian can be written
\begin{equation}
H \,=\, H_1+H_2 \,=\, H_1 \,+\, t_2\!\!\sum_{\;\;\langle i,j\rangle_{{}_{(2)}}}
\!e^{i\,\varphi\,\alpha_{ij}}\,c_i^\dagger\,c_j \,+\, \epsilon_i\,M\,\sum_i\,c_i^\dagger\,c_i \;,
\label{HamHald}
\end{equation}
where $H_1$ is the tight binding Hamiltonian and $H_2$ accounts for the local magnetic fields and for a Semenoff-type parity-breaking term. The first sum in $H_2$ runs on second nearest neighbors sites, $t_2$ being the hopping amplitude, while the second term is the Semenoff contribution coming from on-site potential energy $M$, the prefactor $\epsilon_i=\epsilon_{\pm}=\pm1$ depending on whether the site $i$ is on the first or second sublattice. The Aharonov–Bohm phase $\varphi$ due to the local magnetic fluxes is taken as a parameter of the model and the factor $\alpha_{ij}=\pm1$ gives the chirality of the path related to the second neighbor hopping%
\footnote{it can be derived from the formula
$\alpha_{ij}=\left(\mathbf{v}_i\times\mathbf{v}_j\right)\cdot\mathbf{z}$
where $\mathbf{z}$ is a unit vector orthogonal to the plane and $\mathbf{v}_i$, $\mathbf{v}_j$ are defined as follows: given a pair of second neighbor atoms $i$ and $j$ with a common $k$ first neighbor, $\mathbf{v}_i$, $\mathbf{v}_j$ are the unit vectors along the paths $ik$, $kj$, respectively
}.\par
If we define a basis $\big(\zeta_{\textsc{a}}(\mathbf{k}),\,\zeta_{\textsc{b}}(\mathbf{k})\big)$ of two-component spinors of Bloch states constructed on the two sublattices $\text{A}$ and $\text{B}$, after Fourier-transform the Hamiltonian in the $\mathbf{k}$-space reads:
\begin{equation}
\begin{split}
\mathcal{H}(\mathbf{k})\,=~
    &t_1\,\sum_{i}\,\big(\cos\left(\mathbf{k}\cdot\mathbf{d}_i\right)\,\sigma_1
    +\sin\left(\mathbf{k}\cdot\mathbf{d}_i\right)\,\sigma_2\,\big)
    \,+\,2\;t_2 \cos(\varphi)\,\sum_{i}\,\cos\left(\mathbf{k}\cdot\mathbf{d}^{\mathsmaller{(2)}}_i\right)\,\mathds{1}\;-
    \\[-1ex]
    &-2\;t_2 \sin(\varphi)\,\sum_{i}\,\sin\left(\mathbf{k}\cdot\mathbf{d}^{\mathsmaller{(2)}}_i\right)\,\sigma_3
    \,+\,M\,\sigma_3\;,
\label{HamHaldk}
\end{split}
\end{equation}
where $\sigma_i$ are the Pauli matrices, while $\mathbf{d}_i$ and $\mathbf{d}^{\mathsmaller{(2)}}_i$ are the displacement vectors to the first and second nearest neighbors sites, respectively.
The degeneracy of the bands at the Dirac points is lifted either by non-zero $M$ or non-zero $t_2\sin(\varphi)$, and the fermion masses in the two inequivalent valleys are \cite{Haldane:1988zza}
\begin{equation}
m_i=m_{\pm}=M \mp 3\sqrt{3}\,t_2 \sin{\varphi}\;.
\end{equation}
If we restrict to time-reversal invariant case, $t_2 \sin(\varphi)=0$, the two
masses $m_{+}$ and $m_{-}$ are equal and the system behaves as a Semenoff insulator.\par
The Haldane model can provide an effective description also of specific 2D honeycomb lattices where the time-reversal symmetry is broken without the presence of local magnetic fields \cite{jotzu2014experimental,kim2017realizing}.

\phantomsection 
\addcontentsline{toc}{section}{References}
\bibliographystyle{mybibstyle}
\bibliography{bibliografia}

\providecommand{\href}[2]{#2}\begingroup\raggedright\begin{thebibliography}{10}

\bibitem{Achucarro:1987vz}
A.~Achucarro and P.~K. Townsend, \textit{``{A Chern-Simons Action for
  Three-Dimensional anti-De Sitter Supergravity Theories}''}, Phys. Lett.
  \textbf{B180} (1986) 89. [,732(1987)].

\bibitem{Gaiotto:2008sd}
D.~Gaiotto and E.~Witten, \textit{``{Janus Configurations, Chern-Simons
  Couplings, And The theta-Angle in N=4 Super Yang-Mills Theory}''}, JHEP
  \textbf{06} (2010) 097,
  [\href{http://arxiv.org/abs/0804.2907}{\texttt{arXiv:0804.2907}}].

\bibitem{Kapustin:2009cd}
A.~Kapustin and N.~Saulina, \textit{``{Chern-Simons-Rozansky-Witten topological
  field theory}''}, Nucl. Phys. \textbf{B823} (2009) 403--427,
  [\href{http://arxiv.org/abs/0904.1447}{\texttt{arXiv:0904.1447}}].

\bibitem{Maldacena:1997re}
J.~M. Maldacena, \textit{``{The Large N limit of superconformal field theories
  and supergravity}''}, Int. J. Theor. Phys. \textbf{38} (1999) 1113--1133,
  [\href{http://arxiv.org/abs/hep-th/9711200}{\texttt{hep-th/9711200}}]. [Adv.
  Theor. Math. Phys.2,231(1998)].

\bibitem{Gubser:1998bc}
S.~S. Gubser, I.~R. Klebanov, and A.~M. Polyakov, \textit{``{Gauge theory
  correlators from noncritical string theory}''}, Phys. Lett. \textbf{B428}
  (1998) 105--114,
  [\href{http://arxiv.org/abs/hep-th/9802109}{\texttt{hep-th/9802109}}].

\bibitem{Witten:1998qj}
E.~Witten, \textit{``{Anti-de Sitter space and holography}''}, Adv. Theor.
  Math. Phys. \textbf{2} (1998) 253--291,
  [\href{http://arxiv.org/abs/hep-th/9802150}{\texttt{hep-th/9802150}}].

\bibitem{Alvarez:2011gd}
P.~D. Alvarez, M.~Valenzuela, and J.~Zanelli, \textit{``{Supersymmetry of a
  different kind}''}, JHEP \textbf{04} (2012) 058,
  [\href{http://arxiv.org/abs/1109.3944}{\texttt{arXiv:1109.3944}}].

\bibitem{Guevara:2016rbl}
A.~Guevara, P.~Pais, and J.~Zanelli, \textit{``{Dynamical Contents of
  Unconventional Supersymmetry}''}, JHEP \textbf{08} (2016) 085,
  [\href{http://arxiv.org/abs/1606.0523}{\texttt{arXiv:1606.0523}}].

\bibitem{Iorio:2011yz}
A.~Iorio and G.~Lambiase, \textit{``{The Hawking-Unruh phenomenon on
  graphene}''}, Phys. Lett. \textbf{B716} (2012) 334--337,
  [\href{http://arxiv.org/abs/1108.2340}{\texttt{arXiv:1108.2340}}].

\bibitem{Alvarez:2013tga}
P.~D. Alvarez, P.~Pais, and J.~Zanelli, \textit{``{Unconventional supersymmetry
  and its breaking}''}, Phys. Lett. \textbf{B735} (2014) 314--321,
  [\href{http://arxiv.org/abs/1306.1247}{\texttt{arXiv:1306.1247}}].

\bibitem{Gomes:2017rmd}
Y.~M.~P. Gomes and J.~A. Helayel-Neto, \textit{``{On a five-dimensional
  Chern–Simons AdS supergravity without gravitino}''}, Phys. Lett.
  \textbf{B777} (2018) 275--280,
  [\href{http://arxiv.org/abs/1711.0322}{\texttt{arXiv:1711.0322}}].

\bibitem{Andrianopoli:2018ymh}
L.~Andrianopoli, B.~L. Cerchiai, R.~D'Auria, and M.~Trigiante,
  \textit{``{Unconventional supersymmetry at the boundary of AdS$_{4}$
  supergravity}''}, JHEP \textbf{04} (2018) 007,
  [\href{http://arxiv.org/abs/1801.0808}{\texttt{arXiv:1801.0808}}].

\bibitem{Ezawa:2006cr}
M.~Ezawa, \textit{``{Supersymmetry and unconventional quantum Hall effect in
  graphene}''}, Phys. Lett. \textbf{A372} (2008) 924--929,
  [\href{http://arxiv.org/abs/cond-mat/0606084}{\texttt{cond-mat/0606084}}].

\bibitem{Lee:2006if}
S.-S. Lee, \textit{``{Emergence of supersymmetry at a critical point of a
  lattice model}''}, Phys. Rev. \textbf{B76} (2007) 075103,
  [\href{http://arxiv.org/abs/cond-mat/0611658}{\texttt{cond-mat/0611658}}].

\bibitem{Dartora:2013psa}
C.~A. Dartora and G.~G. Cabrera, \textit{``{Wess-Zumino supersymmetric phase
  and superconductivity in graphene}''}, Phys. Lett. \textbf{A377} (2013)
  907--909.

\bibitem{Andrianopoli:2019sqe}
L.~Andrianopoli, B.~L. Cerchiai, P.~A. Grassi, and M.~Trigiante, \textit{``{The
  Quantum Theory of Chern-Simons Supergravity}''}, JHEP \textbf{06} (2019) 036,
  [\href{http://arxiv.org/abs/1903.0443}{\texttt{arXiv:1903.0443}}].

\bibitem{Iorio:2018agc}
A.~Iorio and P.~Pais, \textit{``{(Anti-)de Sitter, Poincaré, Super symmetries,
  and the two Dirac points of graphene}''}, Annals Phys. \textbf{398} (2018)
  265--286, [\href{http://arxiv.org/abs/1807.0876}{\texttt{arXiv:1807.0876}}].

\bibitem{Andrianopoli:2014aqa}
L.~Andrianopoli and R.~D'Auria, \textit{``{N=1 and N=2 pure supergravities on a
  manifold with boundary}''}, JHEP \textbf{08} (2014) 012,
  [\href{http://arxiv.org/abs/1405.2010}{\texttt{arXiv:1405.2010}}].

\bibitem{Nieh:1981xk}
H.~T. Nieh and M.~L. Yan, \textit{``{Quantized Dirac Field in Curved
  Riemann-cartan Background. 1. Symmetry Properties, Green's Function}''},
  Annals Phys. \textbf{138} (1982) 237.

\bibitem{Chandia:1997hu}
O.~Chandia and J.~Zanelli, \textit{``{Topological invariants, instantons and
  chiral anomaly on spaces with torsion}''}, Phys. Rev. \textbf{D55} (1997)
  7580, [\href{http://arxiv.org/abs/hep-th/9702025}{\texttt{hep-th/9702025}}].

\bibitem{Hughes:2012vg}
T.~L. Hughes, R.~G. Leigh, and O.~Parrikar, \textit{``{Torsional Anomalies,
  Hall Viscosity, and Bulk-boundary Correspondence in Topological States}''},
  Phys. Rev. \textbf{D88} (2013), n.~2, 025040,
  [\href{http://arxiv.org/abs/1211.6442}{\texttt{arXiv:1211.6442}}].

\bibitem{Parrikar:2014usa}
O.~Parrikar, T.~L. Hughes, and R.~G. Leigh, \textit{``{Torsion, Parity-odd
  Response and Anomalies in Topological States}''}, Phys. Rev. \textbf{D90}
  (2014), n.~10, 105004,
  [\href{http://arxiv.org/abs/1407.7043}{\texttt{arXiv:1407.7043}}].

\bibitem{novoselov66}
K.~S. Novoselov, A.~K. Geim, S.~V. Morozov, D.~Jiang, Y.~Zhang, S.~V. Dubonos,
  I.~V. Grigorieva, and A.~A. Firsov, \textit{``Electric field effect in
  atomically thin carbon films''}, Science \textbf{306} (2004), n.~5696,
  666--669.

\bibitem{novoselov2005twodimato}
K.~Novoselov, D.~Jiang, F.~Schedin, T.~Booth, V.~Khotkevich, S.~Morozov, and
  A.~Geim, \textit{``Two-dimensional atomic crystals''}, Proceedings of the
  National Academy of Sciences of the United States of America \textbf{102}
  (2005), n.~30, 10451--10453.

\bibitem{neto2009electronic}
A.~C. Neto, F.~Guinea, N.~M. Peres, K.~S. Novoselov, and A.~K. Geim,
  \textit{``The electronic properties of graphene''}, Reviews of modern physics
  \textbf{81} (2009), n.~1, 109.

\bibitem{katsnelson2007graphene}
M.~Katsnelson and K.~Novoselov, \textit{``Graphene: New bridge between
  condensed matter physics and quantum electrodynamics''}, Solid State
  Communications \textbf{143} (2007), n.~1, 3--13.

\bibitem{Vozmediano:2010zz}
M.~A.~H. Vozmediano, M.~I. Katsnelson, and F.~Guinea, \textit{``{Gauge fields
  in graphene}''}, Phys. Rept. \textbf{496} (2010) 109--148,
  [\href{http://arxiv.org/abs/1003.5179}{\texttt{arXiv:1003.5179}}].

\bibitem{Cortijo:2006xs}
A.~Cortijo and M.~A.~H. Vozmediano, \textit{``{Effects of topological defects
  and local curvature on the electronic properties of planar graphene}''},
  Nucl. Phys. \textbf{B763} (2007) 293--308,
  [\href{http://arxiv.org/abs/cond-mat/0612374}{\texttt{cond-mat/0612374}}].
  [Nucl. Phys.B807,659(2009)].

\bibitem{novoselov2005twodimgas}
K.~Novoselov, A.~K. Geim, S.~Morozov, D.~Jiang, M.~Katsnelson, I.~Grigorieva,
  S.~Dubonos, and A.~Firsov, \textit{``Two-dimensional gas of massless dirac
  fermions in graphene''}, Nature \textbf{438} (2005), n.~7065, 197--200.

\bibitem{Zhang:2005zz}
Y.~Zhang, Y.-W. Tan, H.~L. Stormer, and P.~Kim, \textit{``{Experimental
  observation of the quantum Hall effect and and Berry's phase in graphene}''},
  Nature \textbf{438} (2005) 201--204,
  [\href{http://arxiv.org/abs/cond-mat/0509355}{\texttt{cond-mat/0509355}}].

\bibitem{Gusynin:2006ym}
V.~P. Gusynin, S.~G. Sharapov, and J.~P. Carbotte, \textit{``{Unusual microwave
  response of Dirac quasiparticles in graphene}''}, Phys. Rev. Lett.
  \textbf{96} (2006) 256802,
  [\href{http://arxiv.org/abs/cond-mat/0603267}{\texttt{cond-mat/0603267}}].

\bibitem{Iorio:2013ifa}
A.~Iorio and G.~Lambiase, \textit{``{Quantum field theory in curved graphene
  spacetimes, Lobachevsky geometry, Weyl symmetry, Hawking effect, and all
  that}''}, Phys. Rev. \textbf{D90} (2014), n.~2, 025006,
  [\href{http://arxiv.org/abs/1308.0265}{\texttt{arXiv:1308.0265}}].

\bibitem{geim2007rise}
A.~K. Geim and K.~S. Novoselov, \textit{``The rise of graphene''}, Nature
  materials \textbf{6} (2007), n.~3, 183.

\bibitem{Boada:2010sh}
O.~Boada, A.~Celi, J.~I. Latorre, and M.~Lewenstein, \textit{``{Dirac Equation
  For Cold Atoms In Artificial Curved Spacetimes}''}, New J. Phys. \textbf{13}
  (2011) 035002,
  [\href{http://arxiv.org/abs/1010.1716}{\texttt{arXiv:1010.1716}}].

\bibitem{Iorio:2010pv}
A.~Iorio, \textit{``{Weyl-Gauge Symmetry of Graphene}''}, Annals Phys.
  \textbf{326} (2011) 1334--1353,
  [\href{http://arxiv.org/abs/1007.5012}{\texttt{arXiv:1007.5012}}].

\bibitem{Gallerati:2018dgm}
A.~Gallerati, \textit{``{Graphene properties from curved space Dirac
  equation}''}, Eur. Phys. J. Plus \textbf{134} (2019), n.~5, 202,
  [\href{http://arxiv.org/abs/1808.0118}{\texttt{arXiv:1808.0118}}].

\bibitem{Iorio:2019czs}
M.~F. Ciappina, A.~Iorio, P.~Pais, and A.~Zampeli, \textit{``{Torsion in
  quantum field theory through time-loops on Dirac materials}''},
  \href{http://arxiv.org/abs/1907.0002}{\texttt{arXiv:1907.0002}}.

\bibitem{Semenoff:1984dq}
G.~W. Semenoff, \textit{``{Condensed Matter Simulation of a Three-dimensional
  Anomaly}''}, Phys. Rev. Lett. \textbf{53} (1984) 2449.

\bibitem{Giovannetti2007substrate}
G.~Giovannetti, P.~A. Khomyakov, G.~Brocks, P.~J. Kelly, and J.~Van Den~Brink,
  \textit{``Substrate-induced band gap in graphene on hexagonal boron nitride:
  Ab initio density functional calculations''}, Physical Review B \textbf{76}
  (2007), n.~7, 073103.

\bibitem{zhou2007substrate}
S.~Y. Zhou, G.-H. Gweon, A.~Fedorov, d.~First, PN, W.~De~Heer, D.-H. Lee,
  F.~Guinea, A.~C. Neto, and A.~Lanzara, \textit{``Substrate-induced bandgap
  opening in epitaxial graphene''}, Nature materials \textbf{6} (2007), n.~10,
  770.

\bibitem{Haldane:1988zza}
F.~D.~M. Haldane, \textit{``{Model for a Quantum Hall Effect without Landau
  Levels: Condensed-Matter Realization of the 'Parity Anomaly'}''}, Phys. Rev.
  Lett. \textbf{61} (1988) 2015--2018.

\bibitem{Noris:2019sdw}
R.~Noris and L.~Fatibene, \textit{``{Spin frame transformations and Dirac
  equations}''},
  \href{http://arxiv.org/abs/1910.0463}{\texttt{arXiv:1910.0463}}.

\bibitem{jotzu2014experimental}
G.~Jotzu, M.~Messer, R.~Desbuquois, M.~Lebrat, T.~Uehlinger, D.~Greif, and
  T.~Esslinger, \textit{``Experimental realization of the topological haldane
  model with ultracold fermions''}, Nature \textbf{515} (2014), n.~7526, 237.

\bibitem{kim2017realizing}
H.-S. Kim and H.-Y. Kee, \textit{``Realizing haldane model in fe-based
  honeycomb ferromagnetic insulators''}, npj Quantum Materials \textbf{2}
  (2017), n.~1, 20.

\end{thebibliography}\endgroup

\end{document}